\documentclass[sigconf]{acmart}
\AtBeginDocument{%
  \providecommand\BibTeX{{%
    \normalfont B\kern-0.5em{\scshape i\kern-0.25em b}\kern-0.8em\TeX}}}

\copyrightyear{2019}
\acmYear{2019}
\setcopyright{acmlicensed}
\acmConference[London '19]{London '19: ACM Conference on Computer Communications and Security}{November 11--15, 2019}{London, UK}
\acmPrice{}
\usepackage{graphicx}
\usepackage{epsfig,endnotes}
\usepackage{todonotes}
\usepackage{subfig}
\usepackage{amsmath}
\usepackage{amsfonts}
\usepackage{multirow}
\usepackage{array}
\usepackage{chngcntr}
\DeclareMathOperator*{\argmin}{arg\,min}

\usepackage{chngcntr}

\acmSubmissionID{135}

\begin{document}

%
\title{Learning from Context: Exploiting and Interpreting File Path Information for Better Malware Detection}

\author{Adarsh Kyadige}
\authornote{Both authors contributed equally to this research.}
\email{<first>.<last>@sophos.com}
\orcid{1234-5678-9012}
\author{Ethan M. Rudd}
\authornotemark[1]
\author{Konstantin Berlin}
\affiliation{%
  \institution{Sophos PLC}
  \streetaddress{11730 Plaza America Dr Suite 500}
  \city{Reston}
  \state{Virginia}
  \postcode{20190}
}

%
\renewcommand{\shorttitle}{Learning from Context}

\begin{abstract}

Machine learning (ML) classifiers used for static portable executable (PE) malware detection typically employ single numerical feature vector representations of each file as input with one or more target labels per file during training. However, there is much orthogonal information that can be gleaned from the \textit{context} in which the file was seen. In this paper, we propose utilizing a static source of contextual information -- namely the path of the PE file -- as an auxiliary input to the classifier. While file paths are not malicious or benign in and of themselves, they do provide valuable context for a malicious/benign determination, for example malware droppers may choose file paths to avoid disk scans or select paths that a user is likely to click. Unlike dynamic contextual information, which requires high CPU and runtime overhead, file paths are available with little overhead and can seamlessly be integrated into a multi-view static ML detector, potentially yielding higher detection rates at very high throughput with minimal infrastructural changes. 

Here we propose a multi-view neural network, which takes feature vectors from the PE file content as well as corresponding file paths as inputs and outputs a detection score. To ensure realistic evaluation, we use a commercial-scale dataset of  approximately 10 million samples -- files and file paths from user endpoints of an actual security vendor network. We then conduct an interpretability analysis via LIME modeling to ensure that our classifier has learned a sensible representation and see which parts of the file path most contributed to change in the classifier's score. We find that our model learns useful aspects of the file path for classification, while also learning artifacts from customers testing the vendor's product, e.g., by downloading a directory of malware samples each named as their hash. We prune these artifacts from our test dataset and demonstrate reductions in false negative rate of 32.3\% at a $10^{-3}$ false positive rate (FPR) and 33.1\% at $10^{-4}$ FPR, over a similar topology single input PE file content only model.

\end{abstract}

%
%

%


\maketitle

\section{Introduction}
\label{sec:intro}

Commercial Portable Executable (PE) malware detectors consist of a hybrid of static and dynamic analysis engines. Static detection -- which is fast and effective at detecting a large fraction of malware -- is usually first employed to flag suspicious samples. Static detection involves analyzing the raw PE image on disk and can be performed very quickly, but it is vulnerable to code obfuscation techniques, e.g., compression and polymorphic/metamorphic transformation \cite{moser2007limits}.

Dynamic detection, by contrast, requires running the PE in an emulator and analyzing behavior at run time \cite{egele2012survey}. When dynamic analysis works, it is less susceptible to code obfuscation, but takes substantially greater computational capacity and time to execute than static methods. Moreover, some files are difficult to execute in an emulated environment, but can still be statically analyzed. Consequently, static detection methods are typically the most critical part of an endpoint's malware prevention (blocking malware before it executes) pipeline.

Static detection methods have seen performance advancements recently, thanks to the adoption of machine learning \cite{damodaran2017comparison}, where highly expressive classifiers, e.g., deep neural networks, are fit on labeled data sets of millions of files. When these classifiers are trained, they use \textit{feature vectors} -- numerical descriptions of the static file content -- as input but no auxiliary data. We note, however, that dynamic analysis works well \textit{precisely because of auxiliary data} -- e.g., network traffic, system calls, etc. -- information that cannot be gleaned directly from the static content of the file.  

In this work, we seek to use file paths, as orthogonal input information to augment static ML detectors. File paths are available statically, without any additional instrumentation of the OS, and are already used internally by malware analysts to correct and investigate mischaracterized detections. Using file paths to augment detections on the surface seems potentially problematic, as file paths are not inherently malicious or benign. However, malware droppers often use file paths with certain characteristics for a variety of reasons. For example, a file path may be chosen to increase the likelihood  that a user will execute a malicious PE masquerading as another application, to avoid disk scans, or to hide the files from a user's view. This results in a prevalence of certain types of directory hierarchies, and detectable naming characteristics (e.g., name randomization), which can provide useful hints about the malicious/benign nature of a file, even when this is not immediately obvious from its content. Likewise, file paths corresponding to prevalent types of benignware exhibit certain patterns. By including the file path as an auxiliary input, we are able to combine information about the file, via feature vectors, with information about how likely it is to see such a file in that specific location.

\begin{figure}[!t]
    \begin{center}
\subfloat[File content only (PE) model.]{\includegraphics[width=\linewidth]{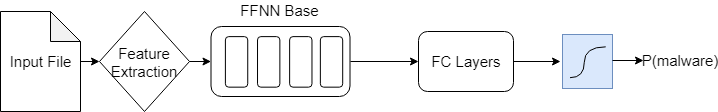}\label{fig:common}}\\
\subfloat[File path content only (FP) model.]{\includegraphics[width=\linewidth]{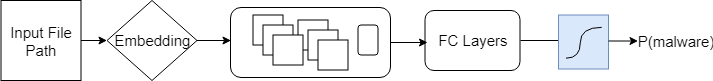}\label{fig:fp_only}}\\
\subfloat[File content + contextual file path (PE + FP) model.]{\includegraphics[width=\linewidth]{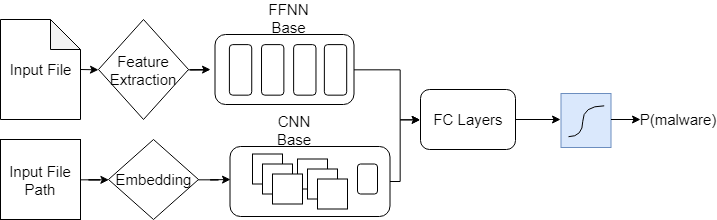}\label{fig:ours}}\\
\caption{Schematic outline of the three approaches that we compare in this paper. \protect\subref{fig:common} A PE content (PE) malware detector, where only static features extracted directly from the PE file are fed to a feed-forward neural network (FFNN). \protect\subref{fig:fp_only} A file path only (FP) malware detector, where the path of each corresponding file is used to determine if that file is malicious or benign. The raw characters are embedded and processed by a series of convolutional layers before being passed to a series of fully connected (FC) layers. \protect\subref{fig:ours} Our novel multi-view approach, where we combine file content with contextual information from the file path (PE+FP). File content features are passed through the same feed-forward neural network base (FFNN Base) layers as in \protect\subref{fig:common} while the file path is passed through the same convolutional neural network base layers (CNN Base) as in \protect\subref{fig:fp_only}. The outputs of these base layers are concatenated together and passed through a series of fully connected layers. Parameters of the separate input paths are jointly optimized. All models output scores between 0 and 1, which represent the confidence of whether a file is malicious.
\label{fig:teaser}}
\end{center}
\end{figure}

We focus our analysis on three models:
\begin{itemize}
    \item The baseline file content only \textit{PE} model, which takes only the PE features as input and outputs a malware confidence score.
    \item Another baseline file path content only \textit{FP} model, which takes only the file's file paths as input and outputs a malware confidence score.
    \item Our proposed multi-view PE file content + contextual file path \textit{PE+FP} model, which takes in both the PE file content features and file paths, and also outputs a malware confidence scores. 
\end{itemize}
A schematic diagram of the three models is shown in Figure \ref{fig:teaser}.

Rather than using vendor aggregation services for our data distribution, which potentially have an artificial file distribution -- i.e., not reflecting a real world deployment case -- and incomplete filepath information, we collect a commercial dataset of actual file and file paths scans on customer endpoints from a large anti-malware vendor, and use them to perform a time split validation of our models.  In addition, we conduct a LIME interpretability analysis \cite{ribeiro2016should} to see what aspects of the file path amplify or attenuate detection on the multi-view model. We find that, while the model learns to detect suspicious aspects of the file path, it also learns to detect artifacts which seem to correspond to vendor's customers performing internal testing the of the product. These artifacts include, e.g., files named by their SHA256 digests, in folders marked ``\textit{malware}'', which were likely bulk-downloaded intentionally. While this is indeed the actual customer distribution, and not data pollution, we do not think detecting vendor tests is an accurate representation of the real world threat landscape. We therefore prune these samples from our test set during evaluation, to avoid presenting a spuriously optimistic view of performance. We find that even after we filter our data, our multi-view classifier trained on both file content and the contextual file path yields statistically significantly better results across the ROC curve and particularly in low false positive rate (FPR) regions. 

The contributions of this paper are as follows:
\begin{enumerate}
    \item We obtain a realistic carefully curated data set of files and file paths from a security vendor's customer endpoints  (rather than a malware / vendor label aggregation service), and carefully prune our test set of ``easy'' samples from customer test endpoints that do not constitute realistic threats in the wild. 

    \item We demonstrate that our multi-view PE+FP malware classifier performs substantially better on our dataset than a model that uses the file contents alone.

    \item We extend Local Interpretable Model Agnostic Explanations (LIME) \cite{ribeiro2016should} to our PE+FP model, and use it to interpret which portions of the file path contribute and detract the most from a detection.

    \item We demonstrate the suitability of PE+FP model as a ranking engine in the context of Endpoint Detection and Response (EDR) applications.

\end{enumerate}

The remainder of this manuscript is structured as follows: Section \ref{sec:background} covers important background concepts and related work. Section \ref{sec:implementation} discusses data set collection and model formulation. Section \ref{sec:experiments} presents an evaluation comparing our novel multi-view approach to a baseline content-only model of similar topology. Section  \ref{sec:discussion} contains a discussion of our results and an interpretability analysis of our model Section \ref{sec:conclusion} concludes. 

\section{Background and Related Work}
\label{sec:background}

In this section, we describe how machine learning is commonly applied to static PE detection and how our approach differs, in a high level sense, by providing contextual information as an auxiliary input. We then present related work in other machine learning domains.

\subsection{Static ML Malware Detection}
\label{sec:static}

Machine learning has been applied in the computer security domain for many years now  \cite{rudd2017survey}, but disruptive performance breakthroughs in static PE models using ML at the commercial scale are a more recent phenomenon. Commercial models typically rely on deep neural networks \cite{saxe2015deep} or boosted decision tree ensembles \cite{anderson2018ember} and have been extended to other static file types as well, including web content \cite{saxe2017expose,saxe2018deep}, office documents \cite{rudd2018meade}, and archives \cite{rudd2018meade}. While methods for dealing with these different input types have their own intricacies, they typically use  single inputs derived from file content as a feature vector or text embedding. 

Static ML detectors use highly parametric classifiers trained on many malicious and benign samples. The goal is to tune the parameters of these classifiers to best match the outputs from the classifiers for all input samples to their actual ground truth labels. Provided that the malware/benignware samples in the training set are similar enough in content to those seen at deployment and that the samples are well labeled, the learned detection function should work well. 

In practice, labels are often collected from vendor aggregation feeds, which submit samples to malware detectors from a variety of vendors. The results can be aggregated into labels that are usually correct, e.g., by using a 1-/5+ criterion \cite{saxe2015deep} or treating the label as a hidden variable and using statistical estimation methods \cite{du2018statistical,kantchelian2015better}. Often a time lag is introduced to let vendors update their models, blacklists, and whitelists accordingly. Generally, the longer the time lag, the more accurate the labels, but the less the data resembles that of the deployment distribution. In actual deployment contexts, classifiers are retrained on new data/labels periodically and the updated parameters are sent to the endpoints on which the detectors are running.

Most static ML for information security (ML-Sec) classifiers operate on learned embeddings over portions of files (e.g., headers) \cite{raff2017learning}, learned embeddings over the full file \cite{raff2018malware}, or most commonly, on pre-engineered numerical \textit{feature vectors} designed to summarize the content from each file \cite{mays2017feature,hassen2017malware,yousefi2017autoencoder,hassen2017scalable,narayanan2016performance,ahmadi2016novel,drew2016polymorphic,saxe2015deep}. Learned embeddings, which generally presume some sort of convolutional architecture, have the advantage that they do not presume a fixed structure and are derived directly during training. However, this process is significantly more expensive, and does not scale as gracefully, e.g., to tens to hundreds of millions of large PE files. Moreover generic bytes are inherently less constrained than inputs like images, video, audio, and text, where convolutions can take advantage of structural localities/heirarchies. Thus, for generic malicious/benign files there is less performance benefit from learning to embed features directly from inputs. Pre-engineered feature vector representations, by contrast quickly distill content from each file that is informative in a classificaiton sense. There are a number of ways to craft feature vectors, including tracking per-byte statistics over sliding windows \cite{ahmadi2016novel,saxe2015deep}, byte histograms\cite{ahmadi2016novel,anderson2018ember}, ngram histograms \cite{mays2017feature}, treating bytes as pixel values in an image (a visualization of the file content) \cite{ahmadi2016novel,mays2017feature}, opcode and function call graph statistics\cite{ahmadi2016novel}, symbol statistics\cite{ahmadi2016novel}, hashed/numerical metadata values \cite{saxe2015deep,anderson2018ember,ahmadi2016novel} -- e.g., entry-point as a fraction of the file, or hashed imports and exports, -- and hashes of delimited tokens \cite{rudd2018meade,drew2016polymorphic}. In practical applications, several different types of feature vectors extracted from file content are often concatenated together to achieve superior performance.

Along a similar vein, our work uses a concatenation of features derived from the content of a PE file as an input to a neural network, but in contradistinction to previous work we add a secondary input which includes contextual information -- namely the PE file path. The PE content input is passed through a series of hidden layers while the file path is passed through a convolutional embedding. Both inputs are ultimately concatenated together into a common  ``stem'' of hidden layers. The final malicious/benign output score is obtained by passing the final dense layer output (a 1-D scalar) through a sigmoid activation function. This is depicted in schematic form in Figure \ref{fig:teaser}.

\subsection{Learning from Multiple Sources} 

Related research in static ML malware detection using deep neural networks has examined learning from multiple sources of information but the approaches are fundamentally different from ours: Huang et al. \cite{huang2016mtnet} and Rudd et al. \cite{rudd2019aloha} use multi-objective learning \cite{caruna1993multitask,rudd2016moon} over multiple auxiliary loss functions which they found increased performance on the main malware detection task. Specifically, Huang et al. introduced an auxiliary categorical cross entropy loss function on mutually exclusive malware family labels, while Rudd et al. introduced several loss functions, including a multi-target binary cross entropy loss over multiple malicious/benign detection sources, a Poisson loss over total detection counts from all malicious/benign sources, and a multi-target binary cross entropy loss over semantic malware attribute tags (e.g., `ransomware', `trojan', `dropper' etc.). While both of these works use multiple target labels derived from metadata about the malicious sample in question, only a single input summarizing the \textit{content} of the sample is provided. Even if the auxiliary labels provide some contextual information to guide the training process, the classification decision itself is still made purely from the content at deployment.  

Our approach utilizes \textit{multiple} input types/modalities -- one which describes the content of the malicious sample, in the form of a PE feature vector similar to \cite{saxe2015deep}, and another which feeds the path of the file to an embedding (similar to \cite{saxe2017expose}) which provides information on where that sample was seen. This technique is a type of \textit{multi-view learning} \cite{xu2013survey}. As the name might suggest, the majority of applications of multi-view learning are in computer vision, where the multiple views \textit{literally} consist of views from  different input cameras/sensors or different views from the same camera/sensor at different times. Early applications were targeted towards detection, localization, and recognition problems \cite{jones2003fast,li2002statistical,wu2004fast}, 2D and 3D modeling and alignment \cite{gross2010multi,blanz2003face,cootes2001active,tola2012efficient,jensen2014large}, and surveillance and tracking \cite{black2002multi}. Later, multi-vew solutions to these problems became popular using deep neural networks \cite{su2015multi,farfade2015multi}. Other common applications of multi-view learning, both in and outside of the computer vision space, include cross-spectral fusion \cite{perera2018in2i}, joint textual/visual content representation for image tagging and retrieval\cite{gong2014multi}, Joint modeling of web page text and inbound hyperlinks \cite{bickel2004multi}, and multi-lingual modeling \cite{faruqui2014improving} to name a few.

As discussed in Section \ref{sec:static}, combining different feature types via concatenation is a common practice in ML-Sec \cite{ronen2018microsoft}, but these approaches -- by and large -- provide different filters on the same content from each file; they do not add contextual information from different input sources. We could only find two approaches in the ML-Sec space which specifically reference themselves as \textit{multi-view}: namely \cite{narayanan2018multi}, in which Narayanan et al. applied multiple kernel learning over dependency graphs for Android malware classification and \cite{bai2016improving}, in which Bai et al. used multi-view ensembles for PE malware detection \cite{bai2016improving}. While these approaches are in some ways similar to ours, they do not use deep learning or contextual information that is exogenous to the malicious/benign files themselves. We are the first, to our knowledge, to perform multi-view modeling for malware detection at commercial data using exogenous file path information fed in conjunction with  file content to a deep neural network.

\section{Implementation Details}
\label{sec:implementation}

In this section we present implementation details of our approach, including the data collection process for obtaining PE files and file paths from customer endpoints, our featurization strategy, and the architectures of our multi-view deep neural network and comparison baselines.

\subsection{Dataset}
\label{sec:implementation:dataset}
For our experiments, we collected training, testing, and validation datasets from a prominent anti-malware vendor's telemetry. This telemetry contains the filepaths and SHA256 digests of portable executable (PE) files seen on their customer endpoints, along with time stamps and other metadata. The telemetry did not contain the raw files due to bandwidth and customer privacy considerations, and instead we used the SHA256 digests to look up and download available files from vendor aggregation services. Malicious/benign labels for these files were computed using a criterion similar to \cite{saxe2015deep,saxe2017expose}, but combined with additional propriety information to generate more accurate labeling. Files that we could not label were removed from the dataset.

We lower-cased all the filepaths for consistency. The file paths that we received from telemetry had drive letters/paths and user names replaced with ``{\texttt{[drive]}}'' and ``{\texttt{[user]}}'' tokens respectively. This step was necessary in order to to protect Potentially Identifiable Information (PII). This obfuscation also has the side benefit of removing near duplicate file paths. We limited the number of file paths associated with each unique PE file sample to a maximum of five first seen paths, in order to avoid ``heavy hitter'' file paths dominating our dataset.

In total, we collected approximately 6 months of sampled telemetry data after performing the above cleaning operations. We split this data into training and test datasets based on the time samples were first seen in our telemetry. Samples that first appeared between June 1 and November 15 2018 were used for training and samples first seen during Jan 1 to Jan 30 2019 were used as a test set. Samples first seen between November 16 and December 1 2018 were used as a validation set to monitor model performance during training, and for model selection and calibration. Care was taken to ensure that there were no overlaps between training, validation, and test sets. The training dataset collected consisted of 9,148,143 distinct samples, with 693,272 of them labeled as malicious. The test dataset had 249,783 total samples with 38,767 of them labeled as malicious. The validation set consisted of 2,225,094 samples with 85,041 of them labeled as malicious.

We note that our original test set contained 275,374 samples. This was reduced to 249,783; by 25,591 samples for the following reason. During an early interpretability analysis using LIME explanations (see Sections \ref{sec:experiments:lime} and \ref{sec:compensating}), we found that a number of files in our test set exhibited particularly high responses with respect to malicious/benign score based off of SHA256 digests in the file path, as well as tokens such as ``malware'', ``prevalent'', etc. Upon investigation, we found that these come from our source vendor's customers (which may include other IT security organizations) testing its endpoint products -- e.g., by downloading folders of malware and seeing if there are resultant detections. While this is, in a sense, indicative of a realistic customer endpoint distribution, in our view, it does not reflect an accurate view of the threat, and including these samples in the test set could yield spuriously optimistic performance evaluations. We therefore pruned our test set of these ``test endpoint'' samples prior to conducting experiments and analysis presented in Section \ref{sec:experiments}. For readers interested in performance comparisons and LIME analysis on the unpruned test set, these results are presented in Appendices \ref{sec:lime_unfiltered} and \ref{sec:perf_unfiltered}.

\subsection{Feature Engineering}

\label{ref:sec_feature_eng}
In order to use file paths in feed-forward neural network, we first needed to convert the variable length strings into numeric vectors of fixed length. We accomplished this using a vectorization scheme similar to \cite{saxe2017expose}, by creating a lookup table keyed on each character with a numeric value (between 0 and the character set size) representing each character. In practice, we implemented this table as a Python dictionary. This transformation required our file paths to be trimmed to a fixed size in order to make it cost effective to perform our experiments. Guided by statistics from our telemetry and early experimentation, we trimmed file paths to the last 100 characters. See Appendix \ref{sec:fp_lengths} for further discussion.

\begin{figure}[!t]
    \centering
    \includegraphics[width=\linewidth]{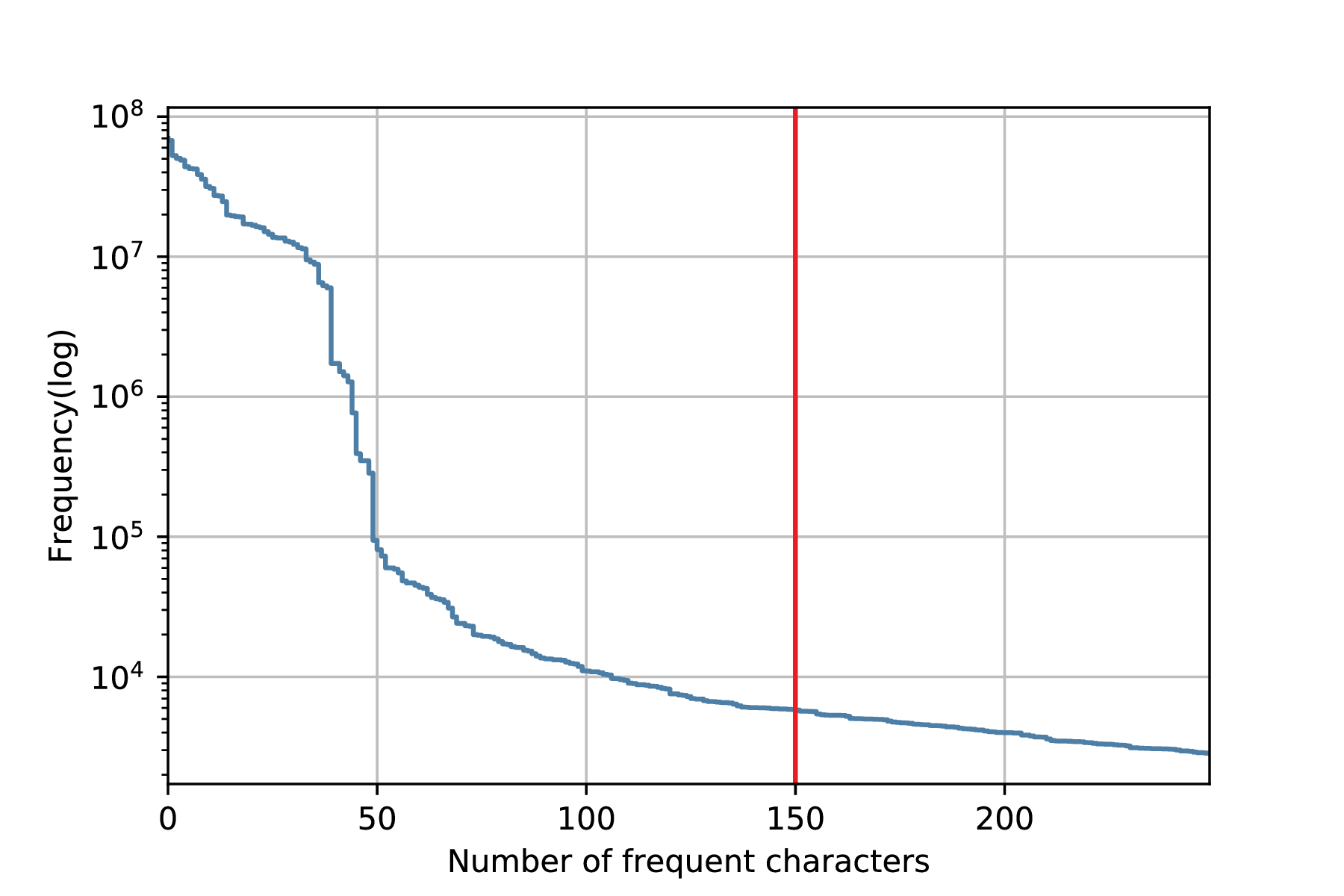}
    \caption{Distribution of Unicode character frequencies by prevalence in our telemetry. The $x$-axis indexes the Unicode character in terms of prevalence rank (most prevalent to least prevalent). The $y$-axis corresponds to frequency. Note the logarithmic scale on the $y$-axis. The red vertical line shows the character rank at which we truncate our vocabulary.}
    \label{fig:vocab_hist}
\end{figure}

In \cite{saxe2017expose}, a character set of 100 printable characters is used as the vocabulary in the lookup table to convert characters to integer vocabulary indices as part of feature construction. In our work, we consider the entire unicode (UTF-8) character set, but limit our vocabulary to 150 most frequently occurring unicode characters, determined by their frequency counts in our data (See Figure \ref{fig:vocab_hist}). We also add a single `other` character that represents all other Unicode characters not in the top 150, and a special null character to represent shorter strings, bringing our vocabulary to a final size of 152.

As features for the content of the PE files, we used floating point 1024-dimensional feature vectors consisting of four distinct feature types, similar to \cite{saxe2015deep}: 
\begin{enumerate}
    \item A 256-dimensional (16x16) 2D histogram of windowed entropy values per byte. A window size of 1024 was selected.
    \item A 256-dimensional (16x16), 2D logarithmically scaled string length/hash histogram.
    \item A 256-dimensional bin of hashes of metadata from the PE header, including PE metadata, including imports, exports, etc.
    \item A 256-dimensional (16x16) byte standard deviation/entropy histogram.
\end{enumerate}

In total, we represent each sample as two feature vectors: a PE content feature vector of 1024 dimensions and a contextual file path feature vector of 100 dimensions.

\subsection{Network Architectures}

\begin{figure*}[!t]
    \centering
    \includegraphics[width=0.8\linewidth]{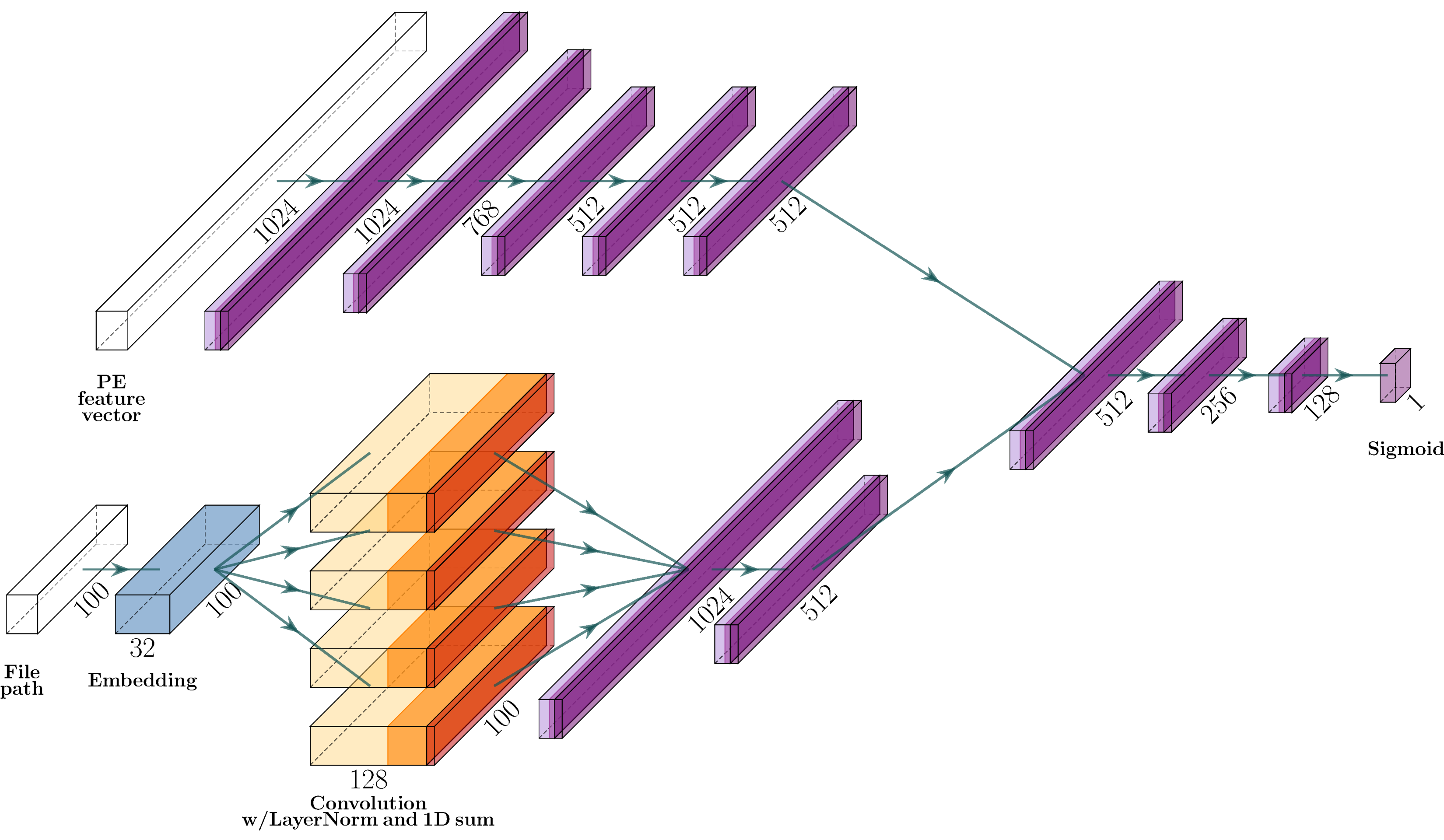}
    \caption{The neural network model we use in our experiments. Each of the unlabeled blocks contains a fully connected layer, followed by Layer Normalization and a Dropout Layer. In experiments where we train the file paths and PE features individually, the respective input and associated input branch is used and the other branch is removed from the model definition.}
    \label{fig:model_diagram}
\end{figure*}

Our multi-view architecture is shown in Figure \ref{fig:model_diagram}. The model has two inputs, the 1024 element PE content feature vector, $\mathbf{x}_{PE}$, and the 100 element file path integer vector, $\mathbf{x}_{FP}$, as described in Section \ref{ref:sec_feature_eng}. Each distinct input is passed through a series of layers with their own parameters, $\theta_{PE}$ and $\theta_{FP}$, for PE features and FP for filepath features respectively, and are jointly optimized during training. The outputs of these layers are then joined (concatenated) and passed through a series of final hidden layers -- a joint output path with parameters $\theta_{O}$. The final output of the network consists of a dense layer followed by a sigmoid activation. Our labeling convention uses $0$ as a benign label and $1$ as a malicious label, so sigmoid outputs close to $1$ are more likely to be malicious than outputs close to $0$, which are more likely to be benign. However, the threshold for malicious/benign determination can be set anywhere along the $(0.0,1.0)$ range according false positive rate (FPR) and detection rate (TPR) tradeoffs for the application at hand -- a reasonable threshold for our use cases is typically at or below $10^{-3}$ FPR.

The PE input arm $\theta_{PE}$ passes $\mathbf{x}_{PE}$ through a series of blocks consisting of four layers each: a Fully Connected layer, a Layer Normalization layer implemented using the technique described in \cite{layernorm}, a Dropout layer with a dropout probability of 0.05, and an Rectified Linear Unit (ReLU) activation. Five of these blocks are connected in sequence with dense layer sizes 1024, 768, 512, 512 and 512 nodes respectively in order. 

The file path input arm $\theta_{FP}$, passes $\mathbf{x}_{FP}$ -- a vector of length 100 -- into an Embedding layer that converts the integer vector into a (100,32) embedding. This embedding is then fed into 4 separate convolution blocks, that contain a 1D convolution layer with 128 filters, a layer normalization layer and a 1D sum layer to flatten the output to a vector. The 4 convolution blocks contain convolution layers with filters of size 2, 3, 4 and 5 respectively that process 2, 3 4 and 5-grams of the input file path. The flattened outputs of these convolution blocks are then concatenated and serve as input to two dense blocks (same form as in the PE input arm). 

The outputs from the fully connected blocks from the PE arm and the file path arm are then concatenated and passed into the joint output path, parameterized by $\theta_O$. This path consists of dense connected blocks (same form as in the PE input arm) of layer sizes 512, 256 and 128. The 128D output of these blocks is then fed to a dense layer which projects the output to 1D, followed by a sigmoid activation that provides the final output of the model.

The PE only model is just the PE+FP model but without the FP arm, taking input $\mathbf{x}_{PE}$ and fitting $\theta_{PE}$ and $\theta_O$ parameters. Similarly, the FP model is the PE+FP model but without the PE arm, taking input $\mathbf{x}_{FP}$ fitting $\theta_{FP}$ and $\theta_O$ paramters. The first layer of the output subnetwork is adjusted appropriately to match the output from the previous layer.

We fit all models using a binary cross entropy loss function. Given the output of our deep learning model $f(\mathbf{x};\theta)$ for input $\mathbf{x}$ with label $y \in \{0,1\}$, and model parameters $\theta$ the loss is:
\begin{equation}
    L(\mathbf{x},y;\theta) = -y \log(f(\mathbf{x};\theta)) + (1-y)\log(1-f(\mathbf{x};\theta)).
\end{equation}

\begin{table*}[!t]
\centering
\caption{Mean and standard deviation true positive rates (TPRs) on the test set for false positive rates (FPRs) of interest. Results were aggregated over five training runs with different weight initializations and minibatch orderings. Best results, shown in \textbf{bold}, consistently occurred when using both feature vectors from the file and contextual file path as inputs. Percentage reduction in mean detection error  in comparison to the PE baseline is shown at the bottom of the table.\label{tab:pruned_results}}
\begin{tabular}{llllll}
\cline{2-6}
& \multicolumn{5}{|c|}{FPR}
\\\multicolumn{1}{l|}{}
&\multicolumn{1}{c|}{$10^{-5}$}
&\multicolumn{1}{c|}{$10^{-4}$}
&\multicolumn{1}{c|}{$10^{-3}$}
&\multicolumn{1}{c|}{$10^{-2}$}
&\multicolumn{1}{c|}{$10^{-1}$}
\\ \hline \hline

\multicolumn{1}{l||}{PE+FP ($0.992 \pm 0.001$ AUC)}
& \multicolumn{1}{l|}{\textbf{0.398} $\pm$ 0.083}
& \multicolumn{1}{l|}{\textbf{0.558} $\pm$ 0.009}
& \multicolumn{1}{l|}{\textbf{0.693} $\pm$ 0.005}
& \multicolumn{1}{l|}{\textbf{0.922} $\pm$ 0.006}
& \multicolumn{1}{l|}{\textbf{0.978} $\pm$ 0.005}
\\\hline

\multicolumn{1}{l||}{PE \,\,\,\,\,\,\,\,\,\,($0.990 \pm 0.002$ AUC)}
& \multicolumn{1}{l|}{0.208 $\pm$ 0.086}
& \multicolumn{1}{l|}{0.339 $\pm$ 0.059}
& \multicolumn{1}{l|}{0.547 $\pm$ 0.007}
& \multicolumn{1}{l|}{0.889 $\pm$ 0.008}
& \multicolumn{1}{l|}{0.972 $\pm$ 0.007}
\\\hline

\multicolumn{1}{l||}{FP \,\,\,\,\,\,\,\,\,\,($0.968 \pm 0.003$ AUC)}
& \multicolumn{1}{l|}{0.02 $\pm$ 0.022}
& \multicolumn{1}{l|}{0.233 $\pm$ 0.04}
& \multicolumn{1}{l|}{0.522 $\pm$ 0.003}
& \multicolumn{1}{l|}{0.711 $\pm$ 0.003}
& \multicolumn{1}{l|}{0.927 $\pm$ 0.003}
\\\hline\hline

\multicolumn{1}{l||}{\% Error Reduction}
& \multicolumn{1}{l|}{24.0}
& \multicolumn{1}{l|}{33.1}
& \multicolumn{1}{l|}{32.3}
& \multicolumn{1}{l|}{30.1}
& \multicolumn{1}{l|}{22.6}
\\ \hline\end{tabular}

\end{table*}

Via an optimizer, we solve for $\hat{\theta}$ the optimal set of parameters that minimize the combined loss over the dataset: 
\begin{equation}
\hat{\theta} = \argmin_{\theta} \sum_{i=1}^{M}L(\mathbf{x}^{(i)},y_i;\theta),
\end{equation}
where $M$ is the number of samples in our dataset, and $y_i$ and $\mathbf{x}^{(i)}$ are the label and the feature vector of the $i$th training sample respectively. 

We built and trained our models using the Keras framework\cite{chollet2015keras}, using the Adam optimizer with Keras's default parameters and $1024$ sized minibatches. Each model is trained for 15 epochs, which we determined was enough for the results to converge.

\section{Experiments and Analysis}
\label{sec:experiments}

We trained three different types of models: two baseline models (PE and FP) and one multi-view model (PE+FP). The baselines can be viewed as different ablations of the multi-view model. One baseline model (PE) takes only PE feature vectors as inputs while the other (FP) takes only file paths as inputs. The multi-view model takes both PE features and file paths as inputs. These model topologies are described in Section \ref{sec:implementation}. We trained each of these models on the same samples; only their inputs differed. The PE baseline is characteristic of a real-world production use case, while the FP baseline should be viewed as a sanity check to ensure that trivial gains do not occur over the PE model by using file path information alone.

To get a statistical view of model performance, we trained five models of each type, with different weight initialization per model, different minibatch ordering, and different seeds for dropout. This allows us to assess not only relative performance comparisons across individual models (as is standard practice), but also mean performance and uncertainty across model types. Training multiple models also tells us important information about the stability of each model type under different initializations. 

\subsection{Performance Evaluation}
\label{sec:performance}

\begin{figure}[ht!]
    \centering
    \includegraphics[width=\linewidth]{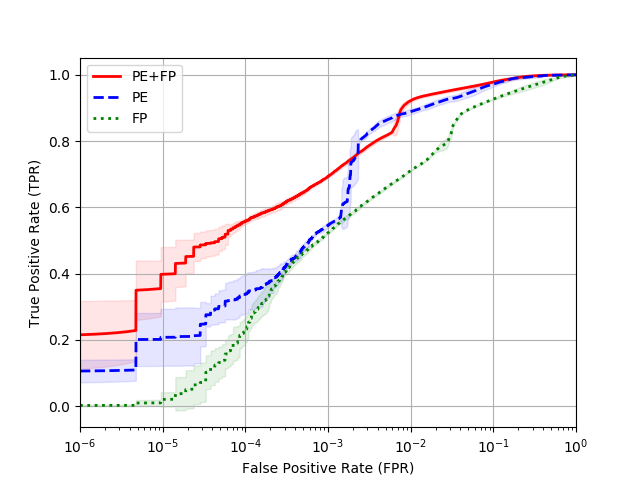}
    \caption{Mean ROC curves and standard deviations for our PE+FP model (red solid line), a PE model (blue dashed line), and an FP model (green dotted line). Mean and uncertainty are computed over five runs.
    }
    \label{fig:pe_fp_feats_roc}
\end{figure}

Results for the three model types, evaluated on the test set -- PE+FP, PE, and FP -- are shown in Figure \ref{fig:pe_fp_feats_roc} as ROC curves and are also summarized in tabular form in in Table \ref{tab:pruned_results}. Recall that these results (mean and standard deviation) were assessed over five runs.

We see that the multi-view (PE+FP) model substantially outperforms the content-only model in terms of net AUC and across the vast majority of the ROC curve, slightly dipping below the PE baseline between $10^{-2}$ and $10^{-3}$ FPR, an effect which could potentially be alleviated with a larger training set. At lower FPRs, the performance improvements from the PE+FP model compared to both baselines is substantial. Specifically, we see that there is a 27\% increase in True Positive rate for the PE + FP model as opposed to the PE model at $10^{-3}$ FPR, and a 64\% increase at $10^{-4}$ FPR. This increase is also accompanied by a reduction in variance of performance, making the PE+FP model a better choice in terms of both stability and overall detection performance. At higher FPR regions, our content-only (PE) model already seems to exhibit very good performance, with a mean TPR of 0.889, and the multi-view (PE + FP) model manages to outperform it, albeit slightly, with a mean TPR of 0.922.  As expected, the filepath only (FP) model that looks only at context consistently performs the worst, with an overall mean AUC of 0.0968, compared to a mean AUC of 0.992 for the multi-view (PE + FP) model and a mean AUC of 0.990 for the content-only (PE) model.  

Note that the TPR/FPR metrics that we use to evaluate detection are invariant to the ratio of malicious to benign samples in our test set. This invariant representation of results is important, since if we are to deploy this model in practice, we can use this TPR/FPR ROC curve to re-calibrate the detector for a a significantly higher ratio of benign to malware by selecting a threshold associated with a low FPR (e.g., $10^{-3}$), rather than the presumed default $0.5$ threshold. It is also for that reason that in our analysis we focus exclusively on the low FPR regions of the curve.

At very low FPRs ($<10^{-4}$) the variance in the TPR increases. This is due to inherent measurement noise at low FPRs: an FPR of $10^{-5}$ means that $1/100,000$ benign samples were falsely labeled as malicious, which is the same order of magnitude as the number of benign samples in our dataset, providing little support for the  numerical interpolation used to generate these ROC curves. Moreover, a small fraction of our test set could potentially be mislabeled. Consequently, results significantly below $10^{-4}$ FPR should be treated with some skepticism. The improvement of the combined model is still substantially larger than the statistical uncertainty for the relevant $10^{-3}$ to $10^{-4}$ FPR regions.

There are two reasons to believe that our test set is more challenging than than real deployment distributions. The first reason is that ML detectors are never deployed by themselves, and are instead guard-railed by signers, prominent file hashes, and AV signature whitelisting. Most of the prominent FP issues can be suppressed using these whitelist approaches. The second reason, is that we removed any previously seen PE file from test set, even it has a new file path. In the raw telemetry, we observed that most executed files are actually not new. However, in our view, the primary job of the ML system is to properly identify previously unseen files, where as old files can typically be whitelisted or blacklisted. Thus, our evaluation reflects the realistic capability of our respective classifiers to detect novel malware.

\subsection{File Path Influence Analysis}
\label{sec:filepath_influence}

\begin{figure*}[!ht]
    \centering
    \includegraphics[width=\linewidth]{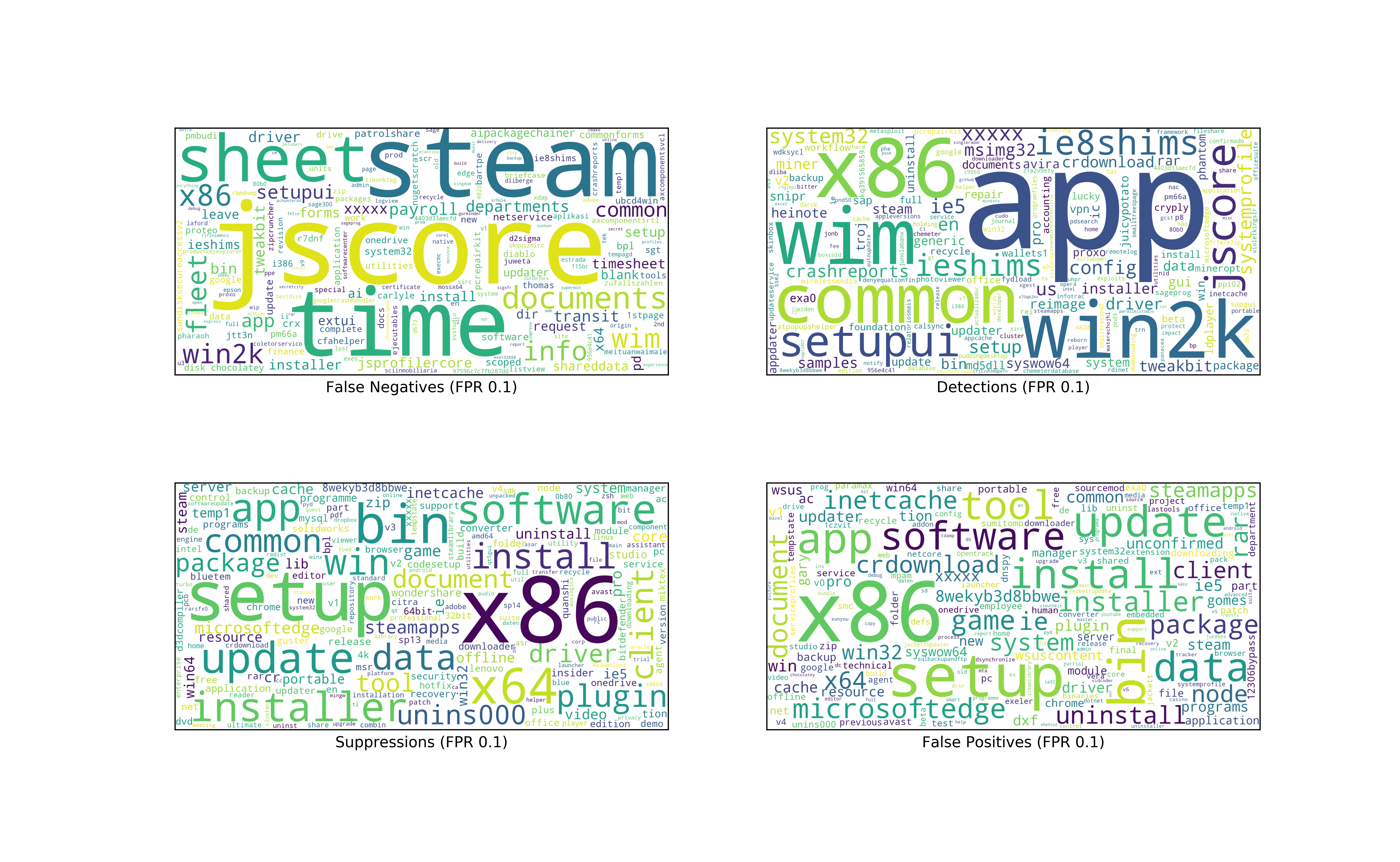}
    \caption{Word clouds generated from file paths where the prediction of the combined PE+FP model is different from the prediction generated by the PE model. These file paths are then divided into four categories based on the initial and changed predictions. In left$\rightarrow$right top$\rightarrow$bottom order, the word clouds are respectively file paths that are additional False Negatives, additional Detections, additional True Negatives, and additional False Positives}
    \label{fig:word_cloud}
\end{figure*}

The overall performance gain from using file paths as additional context information to the neural network model seems evident from the performance metrics in Section \ref{sec:performance}. In this section, we analyze the influence that file paths have on an individual model's performance, by examining additional detections and false negatives introduced by the file paths (PE+FP model) as compared to a model that considers only the PE binary features (PE model). To investigate this, we thresholded detection scores corresponding to particular FPRs and converted continuous model outputs to a binary malware/benign decision. For this analysis, we employed a threshold corresponding to $10^{-1}$ FPR.

A word cloud representation of tokens from the file paths which cause a change in the prediction of the PE+FP model when compared to the PE model is illustrated in Figure \ref{fig:word_cloud}. These file paths are represented by four word clouds, one each for additional false negatives, additional detections, additional true negatives, and additional false positives, to get a representation that captures the most frequent kinds of tokens in file paths that were detected or misclassified. Common file roots such as ``\small{\texttt{program files}}'', ``\small{\texttt{appdata}}'', etc were filtered out as stop words in order to avoid these words crowding the analysis and suppressing more interesting patterns.

Looking at the word cloud for additional detections by the combined model, we observed that \small{\texttt{app}} occurred very frequently. Upon further inspection, we discovered a family of Trojans that always had a file path of the pattern: ``\small{\texttt{\lbrack drive\rbrack\textbackslash Users\textbackslash \lbrack user\rbrack\textbackslash appdata\textbackslash local\textbackslash temp\textbackslash \\ \lbrack random  9  digit  sequence\rbrack\textbackslash app.exe}}'', with about 10000 such occurrences in the training set. Out of the 969 such occurrences in the test set, we observed that the PE features only model detected just 152 variants, whereas the PE + file path model detected an additional 575 samples as malicious (at an FPR of 0.001). This improvement in detections at a low FPR is very encouraging, and in line with the intuition for using file paths. We observe several such occurrences that cause successful detections because of patterns that malicious files exhibit in their file paths, with executables residing in common Windows folders such as \small{\texttt{\%SYSTEM\%}}, \small{\texttt{\%USERPROFILE\%}}, \small{\texttt{\%APPDATA\%}}, \small{\texttt{\%SYSTEMPROFILE\%}}, etc.

However, this does not mean that our model relies only on the file path and makes trivial predictions. Since file paths are not used individually, but as context along with PE content, the model seems to learn to dynamically attribute value to different parts of the file path based on specific patterns in file content that are unique to those files. In other words, there is no single file path that can always cause a detection or a suppression regardless of what file content is associated with it. This is clear when we look at a set of malicious samples in our data that seem to impersonate unfinished chrome downloads on disk. These files are of the pattern ``\small{\texttt{\lbrack drive\rbrack \textbackslash Users\textbackslash \lbrack user\rbrack \textbackslash Downloads\textbackslash Unconfirmed\textbackslash\\ \lbrack random 6 digit sequence\rbrack.crdownload}}''. Since both malicious and benign files have similar file path patterns in this case, the detection rates for both the combined (PE + FP) model and the content only (PE) model remain virtually the same, signaling that the file path has almost no influence on the prediction in this case.

While introducing file paths as an auxiliary input yields a compelling performance improvement, in some scenarios, it also causes misclassifications. Based on the analysis of missed detections and false negatives from PE+FP model, we observed the following modes of failure. 

\begin{itemize}
    \item Malicious files contained in system restore checkpoints and deleted files in the recycle bin usually have completely randomly generated filenames, with a very large percentage of them being benign. We have observed that malicious files convicted with a low confidence by the PE model are sometimes marked as benign by the PE+FP model.
    \item Novel malicious files with names associated mostly with benign files in the training set are also marked falsely as benign files. This does not happen very often. The problem is most chronic in cases when a large set of benign files with similar file paths are seen in the training set, and the detection confidence from the PE features is low. For example, in the False Negatives word cloud in Figure \ref{fig:word_cloud}, we see that file paths containing the words ``\small{\texttt{time sheet}}'', ``\small{\texttt{steam}}'', ``\small{\texttt{fleet info}}'', ``\small{\texttt{payroll}}'', ``\small{\texttt{departments}}'', etc are wrongly exonerated, because these names are largely associated with benign files.
\end{itemize}

\begin{table}[!th]
\centering
\caption{Number of samples where our model has additional Detections and additional False Negatives as opposed to a model using just the PE binary features, at different FPR levels.\label{tab:net_gain}}
\begin{tabular}{llllll}
\cline{2-6}

& \multicolumn{5}{|c|}{FPR}
\\\multicolumn{1}{l|}{}
&\multicolumn{1}{c|}{$10^{-5}$}
&\multicolumn{1}{c|}{$10^{-4}$}
&\multicolumn{1}{c|}{$10^{-3}$}
&\multicolumn{1}{c|}{$10^{-2}$}
&\multicolumn{1}{c|}{$10^{-1}$}
\\ \hline \hline

\multicolumn{1}{l||}{Additional TPs}
& \multicolumn{1}{l|}{7796}
& \multicolumn{1}{l|}{9099}
& \multicolumn{1}{l|}{3601}
& \multicolumn{1}{l|}{2229}
& \multicolumn{1}{l|}{734}
\\\hline

\multicolumn{1}{l||}{Additional FNs}
& \multicolumn{1}{l|}{90}
& \multicolumn{1}{l|}{475}
& \multicolumn{1}{l|}{1799}
& \multicolumn{1}{l|}{565}
& \multicolumn{1}{l|}{310}
\\\hline

\multicolumn{1}{l||}{Net Gain}
& \multicolumn{1}{l|}{7706}
& \multicolumn{1}{l|}{8624}
& \multicolumn{1}{l|}{1802}
& \multicolumn{1}{l|}{1664}
& \multicolumn{1}{l|}{424}
\\\hline\end{tabular}

\end{table}

Fortunately, the occurrence of such failures seems to be quite low compared to the number of files we are able to convict using file path information. These failure modes also happen only when the PE model is almost completely ambivalent about its prediction. It is in this set of gray files that the PE+FP model produces additional detections and suppresses false positives, while occasionally missing a few detections. This is especially impressive when considering that every sample in the test set is completely unseen in the training distribution. Table \ref{tab:net_gain} shows the net detection gain at different FPR levels for the PE+FP model over the PE model. 

We also see that the word clouds for additional true negatives and false positives are almost identical. Since we are controlling for a fixed FPR  while generating the word clouds, it is important to note that the number of additional true negatives is generally equal to the number of false positives. However, the distribution changes, albeit slightly. From manual inspection, we found that most files whose predictions changed to  negatives/false positives were generally close to the decision threshold with randomly generated components in their file paths, which likely cause minor changes in predicted probabilities and the tendency to hop between predictions.

\subsection{LIME Analysis}
\label{sec:experiments:lime}

To ensure that our multi-view model has learned meaningful content from PE file paths, we pick one of our trained models and employ Local Interpretable Model-Agnostic Explanations (LIME) introduced by Ribiero et al. in \cite{ribeiro2016should} to samples from the test set. LIME explanations assume that a trained ML model, in our case $f(\mathbf{x}; \hat\theta)$, can be explained by a simple interpretable linear model locally, around an input $\mathbf{x}$. Based on this assumption, which the authors justify in \cite{ribeiro2016should}, the linear model is trained to approximate $f(\mathbf{x}; \hat{\theta})$ within a neighborhood of $\mathbf{x}$. The learned model weights are then used to judge feature importance of the original deep model.

The definition of a realistic neighborhood around a specific input $\mathbf{x}$ is problem specific, and represents the main challenge in adapting LIME to our file path analysis. We generated the neighborhood samples, by first tokenizing the file path by five delimeters: ``\textbackslash'', ``/'', ``.'', ``\_'', and ``-''. We then selected one random token to perturb. We crafted our perturbations by first sampling a random number from the uniform distribution on $\mathbb{R} \in [0.0,1.0]$. If the number was greater than 0.5, we replaced the token with a random string; if it was less than or equal to 0.5, we removed the token and preceding delimiter.

After applying our perturbations, we then one-hot encoded all the possible tokens into a feature vector $\mathbf{v}$. We fit each LIME classifier on $N=5000$ such perturbations, labeling the original sample $\mathbf{v}^{(0)}$, and the rest as $\mathbf{v}^{(1)} \ldots  \mathbf{v}^{(N)}$. Similarly, we reconstituted, the modified strings $\mathbf{x}^{(1)} \ldots  \mathbf{x}^{(N)}$, by recombining separators and tokens back into a full string. 

As an example, consider the following file path:
\begin{center}
\small{\texttt{C:\textbackslash users\textbackslash Bob\textbackslash appdata\textbackslash local\textbackslash temp\textbackslash rar\textbackslash payment.scr}}.
\end{center}
This pre-processed file path, after substituting the drive and user name, will be  as follows:
\begin{center}
\small\texttt{[drive]\textbackslash users\textbackslash [user]\textbackslash appdata\textbackslash local\textbackslash temp\textbackslash rar\textbackslash payment.scr}.
\end{center}
Splitting on delimiters ``\textbackslash'' and ``.'' yields $9$ distinct tokens. We randomly generate three more samples, which creates two new tokens, resulting in $11$ distinct tokens in our example dataset:
\setcounter{MaxMatrixCols}{20}
\begin{align*}
    \mathbf{v}^{(1)} &= \begin{bmatrix} 1 & 1 & 1 & 1 & 0 & 1 & 1 & 1 & 1 & 0 & 0 \end{bmatrix}, \\
    \mathbf{v}^{(2)} &= \begin{bmatrix} 1 & 1 & 1 & 0 & 1 & 1 & 1 & 1 & 1 & 1 & 0 \end{bmatrix},\\
    \mathbf{v}^{(3)} &= \begin{bmatrix} 1 & 1 & 1 & 0 & 1 & 1 & 1 & 1 & 1 & 0 & 1 \end{bmatrix},
\end{align*}
where 
\begin{equation*}
    \mathbf{v}^{(0)} = \begin{bmatrix} 1 & 1 & 1 & 1 & 1 & 1 & 1 & 1 & 1 & 0 & 0 \end{bmatrix}.
\end{equation*}

Corresponding perturbations generated by our perturbation routine appear as follows: 
\small{
\begin{align*}
    \mathbf{x}^{(1)} &= \text{\texttt{[drive]\textbackslash users\textbackslash [user]\textbackslash appdata\textbackslash temp\textbackslash rar\textbackslash payment.scr}}\\
    \mathbf{x}^{(2)} &= \text{\texttt{[drive]\textbackslash users\textbackslash [user]\textbackslash ztakzfw\textbackslash local\textbackslash temp\textbackslash rar\textbackslash payment.scr}},\\
    \mathbf{x}^{(3)} &= \text{\texttt{[drive]\textbackslash users\textbackslash [user]\textbackslash nhus11n\textbackslash local\textbackslash temp\textbackslash rar\textbackslash payment.scr}}.\\
\end{align*}
}
In our example $\mathbf{x}^{(1)}_5$, corresponding to token ``local'' has a value of $0$, since it no longer exists in the string. In $\mathbf{x}^{(2)}$, we perturb the delimiter corresponding to ``appdata'', replacing it by a random string and thus generating a token that did not exist in the original sample. We do this again for $\mathbf{x}^{(3)}$ with another random string.

Consistent with Ribiero et al., we fit our model using Lasso regression -- least squares regression with an $L_1$ penalty, which has the effect of encouraging sparsity in the explanation. The overall optimization objective of the LIME explanation model is:

\begin{equation}
\argmin_{\phi} \sum_{i=0}^{N} w_i \left(\phi^T \mathbf{v}^{(i)}   - f(\mathbf{x}^{(i)}; \hat\theta)\right)^2 + \lambda ||\phi||_1, 
\end{equation}

\noindent
where $\phi$ are the parameter weights of the LIME model, $\lambda$ is the weight regularization penalty, and $w_i$ is the weight associated with the $i$th sample. For our implementation, we used Scikit-learn's \cite{pedregosa2011scikit} default Lasso regression implementation, with $\lambda$ value of $0.01$, which we observed induce a sparse solution. For numerical stability, here we take $f(\cdot)$ to be the prediction of the network prior to the sigmoid output.

Ribiero et al. recommend computing $\mathbf{w}$ using distance kernel from the original sample to the target sample. This is enforce locality so that perturbations closer to the original sample contribute more to the regression objective. For our formulation, all perturbations have approximately the same semantic distance, so set $w_1\ldots w_N$ to $1.0$, and  $w_0=\sum_{j=1}^N w_j$, in order to give enough weight to the one original sample so the prediction approximately matches the original $f(\mathbf{x}; \hat\theta)$ values after fitting.

\begin{figure*}[!h]
    \subfloat[Positive (Increase)]{\includegraphics[width=0.75\textwidth]{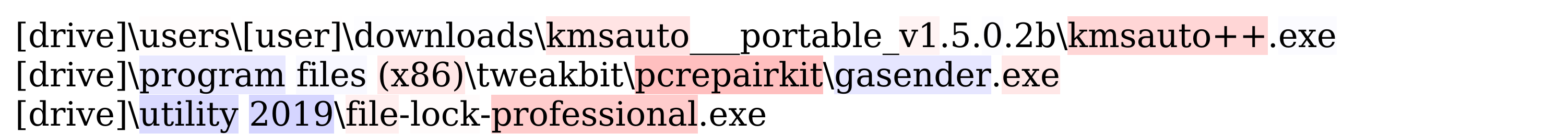}\label{fig:lime_examples_pos}}\\
    \subfloat[Negative (Decrease)]{\includegraphics[width=0.75\textwidth]{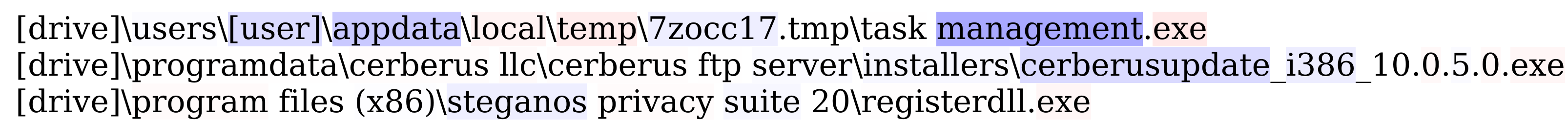}\label{fig:lime_examples_neg}}
    \caption{Example file paths from our LIME analysis with \protect\subref{fig:lime_examples_pos} positive and \protect\subref{fig:lime_examples_neg} negative ground truth labels. The path tokens are highlighted based on the Lasso weights, as computed by the LIME model. As the model is linear, the token weights can be directly interpreted as either making the overall malware score higher (red weights) or lower (blue weights). The color shading is proportional to weight amplitude, i.e.,  darker red and blue shades correspond to greater magnitude weights, while lighter shades correspond to smaller magnitudes. White color corresponds to no impact.}
    \label{fig:Posandnegurls}
\end{figure*}

We visualized the computed Lasso model weights for several interesting examples in Figure \ref{fig:Posandnegurls}, by overlaying the computed weights on top of the file path string.

 In the first positive example we can see that that the token ``\texttt{kmsauto}'' is being identified as a maliciousness indicator by our PE+FP model. KMS Auto is legally dubious Microsoft product activator, and this file is identified as ``PUA:Win32/AutoKMS'' by Microsoft. Similar, in the second second positive example our PE+FP model gave high score to ``\texttt{pcrepairkit}''. Repair kits are typically questionable software products that usually contain spyware or malware.
 
 On the other hand, in the several negative examples we can see that management tools are being down-weighted by the PE+FP model, as compared to the PE model. Management tools are notoriously difficult to distinguish from spyware, as their functionality is basically the same, the only difference is intent of the user. In this case using filepath information provided us more context for the detection, thus allowing more accurate identification by the PE+FP model. We note that these are a few interesting examples, and that the relative contributions of tokens also have a non-linear dependency on the file content itself. For example, when we kept the same path for the first negative example, but replaced the file with a randomly chosen malicious file, the importance of the token ``\texttt{management}'' was significantly reduced.

Finally, we performed an aggregate LIME analysis to identify prominent tokens throughout our dataset, by choosing 200 samples from our test data set to analyze -- 100 with a malicious ground truth label and 100 with a benign ground truth label. The first 100 of these samples consisted of positive ground truth label test samples where the score from the PE+FP model most significantly increased beyond that of PE baseline. The second 100 consisted of negative ground truth label samples where the score of the PE+FP model most significantly decreased beneath the score of the PE baseline. Note that measuring \textit{most significant increase} and \textit{most significant decrease}, on raw output scores is potentially problematic because different models -- even ones with bounded sigmoid outputs -- may have fundamentally different score scales between 0 and 1. Therefore, we performed calibration via isotonic regression on scores over the validation set for each model before assessing score differences. We then aggregated LIME parameter weights across tokens ($\theta_{ij}$) and normalized by token frequency, looking at tokens of highest and lowest weights for the selected 200 samples. The top 10 tokens which increased and decreased response are shown in Table \ref{tab:tokens_weights_pruned}.

\begin{table*}[!t]

\begin{center}
\caption{Tokens and corresponding weights from our LIME analysis that most amplified and attenuated responses for malicious and benign samples. For the malicious samples analyzed, tokens that resulted in greatest increase and greatest decrease in classification score are shown in  \protect\subref{tab:inc_mal} and \protect\subref{tab:dec_mal}. Corresponding tokens for the benign samples are shown in  \protect\subref{tab:inc_ben} and \protect\subref{tab:dec_ben}. Malicious samples were selected from the 100 samples from the test data set where the ground truth label was malicious and the calibrated score from the PE+File Path model increased the most above the calibrated score from the PE model. Benign samples were similarly selected from the pruned data set  according to the most significant drop in calibrated score.\label{tab:tokens_weights_pruned}}

\subfloat[Increase (Malicious)]{
\begin{tabular}{lll}
\textbf{Token} & \textbf{Weight}\\
\hline
\url{2786} & 7.436\\
\url{4327} & 5.854\\
\url{8o0sdtwhrxkz} & 4.213\\
\url{28pygyuokzwwn} & 3.826\\
\url{wfzctyetugjwxxuy} & 3.736\\
\url{3015798005} & 3.592\\
\url{setup} & 3.313\\
\url{jzljumnkfaapzpqq} & 3.183\\
\url{whyovxk3mplt6} & 3.167\\
\url{1467} & 2.219\\
\end{tabular}
\label{tab:inc_mal}}
\subfloat[Decrease (Malicious)]{

\begin{tabular}{lll}
\textbf{Token} & \textbf{Weight}\\
\hline
\url{onv2k} & -6.677\\
\url{computerz} & -6.433\\
\url{westlake} & -5.565\\
\url{editor} & -5.13\\
\url{printingtools} & -4.738\\
\url{videodecodesdk} & -3.687\\
\url{placar80} & -3.663\\
\url{movavistatistics} & -3.556\\
\url{enterprise} & -3.488\\
\url{jarvee} & -3.401\\
\end{tabular}
\label{tab:dec_mal}
}
\subfloat[Increase (Benign)]{
\begin{tabular}{lll}
\textbf{Token} & \textbf{Weight}\\
\hline
\url{miner} & 9.369\\
\url{z} & 8.163\\
\url{2639} & 6.876\\
\url{mineropt} & 6.507\\
\url{2198205786} & 6.28\\
\url{systemprofile} & 4.26\\
\url{xxxxx} & 4.193\\
\url{t} & 3.916\\
\url{d} & 3.812\\
\url{namespace} & 3.441\\
\end{tabular}
\label{tab:inc_ben}
}
\subfloat[Decrease (Benign)]{
\begin{tabular}{lll}
\textbf{Token} & \textbf{Weight}\\
\hline
\url{msi61f0} & -8.04\\
\url{part} & -7.022\\
\url{ciscosparklauncher} & -6.642\\
\url{sesinaci} & -4.738\\
\url{clientinst} & -4.445\\
\url{safesenderslist} & -4.443\\
\url{setup} & -4.389\\
\url{sd} & -4.147\\
\url{wim} & -4.06\\
\url{ie8shims} & -3.996\\
\end{tabular}
\label{tab:dec_ben}
}
\end{center}

\end{table*}

The results of running our LIME analysis, are shown in Table \ref{tab:tokens_weights_pruned}. For malicious samples, we see that the tokens of highest weight consisted of strings with randomized content, that were not cryptographic digests, perhaps an attempt at obfuscation. The remaining high-weight token, \small{\texttt{setup}} is perhaps indicative of an infected installer. Tokens with large negative weights consist of common looking benign software names, as one might expect. Of the benign samples that we assessed, tokens that increased response tended to have very short length, e.g., ``\small{\texttt{t}}'', ``\small{\texttt{d}}'', and ``\small{\texttt{z}}'', very high or very low entropy, e.g., ``\small{\texttt{219805786}}'' and ``\small{\texttt{xxxxx}}'', and have ``miner'' in their names, e.g., ``\small{\texttt{miner}}'', ``\small{\texttt{mineropt}}'' -- indicating the likely presence of a (benign) cryptocurrency miner, potentially downloaded by the user voluntarily. It is not surprising that the string ``miner'' increased response as many types of malware and potentially unwanted benignware steal CPU cycles to mine cryptocurrency. With respect to tokens that most attenuated the response, they appear to be components of standard software. Interestingly, ``\small{\texttt{setup}}'' tends to attenuate response for the benignware that we analyzed, indicating that the behavior of tokens depends on their contexutal location within the file path. Note that, as LIME involves fitting a classifier per sample, this analysis is limited only to the samples that we analyzed. However, it suggests that our neural network is learning to extract useful contextual information from file paths; not just mere data artifacts.

\subsection{Model Debugging with LIME}
\label{sec:compensating}

As seen in Section \ref{sec:experiments:lime} above, LIME can serve to interpret which parts of test samples triggered a high or low classification response. In this section, we highlight the ability of lime to help debug overfitting in models and ensure that seemingly optimistic test results are not driven by spurious correlations. 

During our initial experiments, when we performed LIME analysis on a trained PE + FP model, we observed some interesting patterns among tokens with a disproportionately high response. The tokens that triggered the greatest increase were all hexadecimal digests that seemed to be associated with malicious PE files named with the SHA256 digest of their contents. These LIME results are presented in Table \ref{tab:tokens_weights} in the Appendix.
 
We concluded that these high-response tokens come from customers who were intentionally testing the detection capability of the source vendor from which we obtained our data set. This is likely a standard and fairly common occurrence, as the easiest\footnote{We do not think this is actually a good way to test vendor efficacy, as the test can clearly be gamed by a file path model such as ours, as well as in the cloud blacklists.} way to test an anti-virus engine is to simply download known malware datasets, and see if they are detected. While this type of data is a microcosm of malware and benignware ``in-the-wild'' distributions, it is not representative of realistic threats and using this data in our analysis could lead to an overly optimistic measure of FP and PE+FP performance.
 
Therefore, as we mentioned in Section \ref{sec:implementation:dataset}, we removed all samples from the test set with file paths containing the names ``\small{\texttt{malware}}'', ``\small{\texttt{prevalent}}'', and our source organization's name, as well as strings of length 10 or greater with hexadecimal-only characters, corresponding to hash digests and re-ran our evaluation. This reduced the size of our test set from 275374 to 249783; by 25591 samples.

Comparative model performances on the unpruned test set are shown in ROCs in Figure \ref{fig:pruned_unpruned} of the appendix and in tabular form in Table \ref{tab:results}. Figure \ref{fig:pruned_unpruned} demonstrates the difference in model performance before and after dataset filtering. As expected, we see slightly better performance from the PE+FP model on the unpruned data set, since all the easy-to-classify file paths are included in this dataset. The performance of the content-only (PE) model is largely unchanged by pruning, while the performance of the file path (FP) model is diminished by pruning. We present this result to demonstrate the importance of selecting a meaningful/representative test set, particularly when dealing with multiple input types, and also to highlight the utility of LIME in model debugging.

\section{Discussion}
\label{sec:discussion}

In this section, we discuss practical applications and potential issues of associated with deploying our multi-view model. First, in Section \ref{sec:adversarial}, we explore the vulnerability of our model in an adversarial setting. Then in Section \ref{sec:edr} we explore the utility of our model in an endpoint detection and response (EDR) context.

\subsection{Sensitivity to Adversarial Attacks}
\label{sec:adversarial}

A natural concern when using file paths for static detection is that an attacker has a fair bit of control over where on the system malware can reside. This adds another input which an adversary can manipulate to evade detection -- much as a PE content-only model can be evaded, e.g., by adding overlays from benign software, a PE+FP model can potentially be evaded by using common paths from benign files \textit{and/or} by modifying PE content. Defense/hardening against adversarial attacks is an active area of research in the anti-malware and ML-Sec communities, which we consider addressing beyond the scope of this paper. However, it is important to be aware of the potential issue that adversarial attacks pose for static ML detectors and particularly for our model.

In practice, all deployed static ML detectors that we are aware of have some susceptibility to adversarial attacks. However, most vendor products contain a variety of detection methods -- both ML-based and non-ML-based, somewhat ameliorating the threat. In these contexts, the role of static ML detectors is to serve as an accurate filter that catches malware at scale in a manner independent of other detection components in the anti-malware stack-- not necessarily as a catch-all solution. At the time of this writing we are also unaware of widespread adversarial attacks in the wild. Thus, while the potential sensitivity to adversarial attacks is important to acknowledge, this does not preclude using our model for many production applications. Finally, the practical nature of altering file location of a piece of malware could inherently diminish its effectiveness: as discussed in Section \ref{sec:intro}, file paths which provide the most useful information are often specifically chosen for a reason -- e.g., to evade disk scans or to increase the likelihood that a user will open an infected application. Thus an adversary might \textit{become less effective} in a malware campaign by altering these chosen file paths. 

\subsection{Applications in Endpoint Detection and Response (EDR)}
\label{sec:edr}

Some organizations might consider deploying the PE+FP model on an endpoint in blocking mode as high-risk due to potentially unexpected behavior and concerns about adversarial manipulations, 

In these cases, the PE+FP model can be moved to the backend, and used in an Endpoint Detection and Response (EDR) context, by augmenting Security Operations Center (SOC) team's threat hunting operations. Here our PE+FP model can be used as a secondary model to flag samples that registered below the FPR threshold of the deployment model as suspicious for inspection by human analysts. In these settings there is a budget in terms of samples that human analysts can inspect (manual detection is time consuming). 
\begin{figure}[!t]
    \centering
    \includegraphics[width=\linewidth]{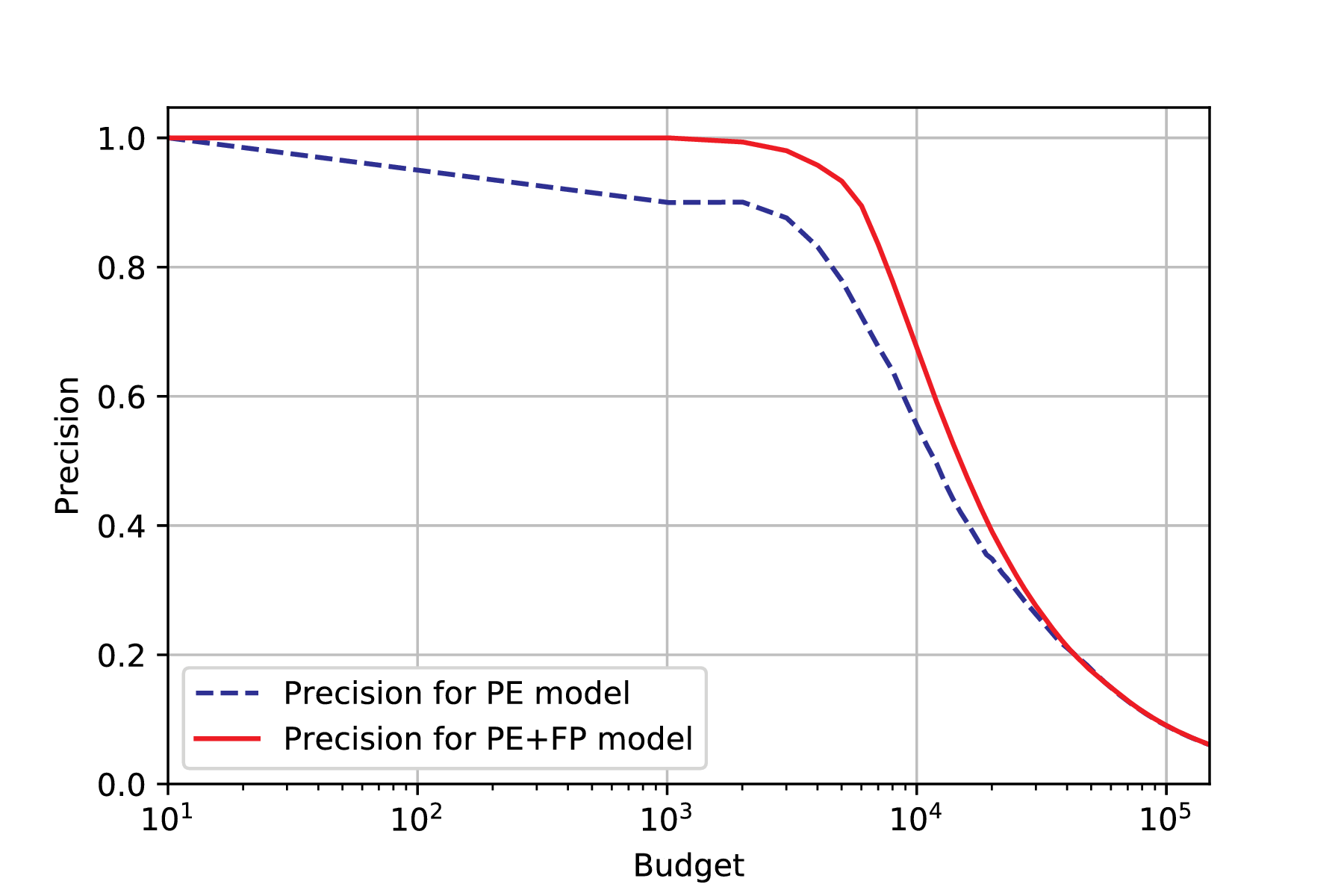}
    \caption{
        Precision curves for the PE only model and the multi-view model as a function of budget, ranging from $10$ to $50000$ samples. PE only model precision is shown using the blue dotted line and the multi-view model precision is shown using the red dashed line. Note that the $x$-axis uses a logarithmic scale
    }
    \label{fig:edr}
\end{figure}

To evaluate the suitability of our model in an EDR setting, we use the following methodology: We threshold a trained PE model at an FPR of $10^{-3}$, which reflects a typical deployment use case, convicting samples which it is highly certain about as malicious, i.e., any event for which the model output is higher than the threshold is considered a certain malicious event. Excluding these convicted samples from our analysis, we then demonstrate how the PE+FP model can improve precision in retrieving malicious events for budgets ranging from 10 up to 50000 events, where the budget is defined as how many samples can be manually inspected by humans. Figure \ref{fig:edr} shows a performance comparison of the precision at different budgets, when using the PE only model and the multi-view model in an EDR use case. 

It is clear from the plot that the multi-view model retrieves significantly more malicious events for a given budget, especially for lower budgets. This suggests that our multi-view model could be used effectively in an EDR application that does not require a large change in existing static ML deployments, but yields significant detection gains during threat hunting.

\section{Conclusion}
\label{sec:conclusion}

In this paper we have demonstrated that deep neural network malware detectors can benefit from contextual information from file paths, even when this information is not inherently malicious or benign. Adding file paths to our detection model did not require any additional endpoint instrumentation, and provided a statistically significant improvement in the overall ROC curve, throughout relevant FPR regions. The fact that we measured the performance of our models directly on a customer endpoint distribution suggests that our multi-view model can practically be deployed to endpoints, even though there are some logistical and user interface issues that might need to be addressed, since moving a file between directories could change the detection scores.

One potentially powerful spin-off of our multi-view detection approach would be to use it in a behavioral detection engine, i.e., use static features along with behavioral data as  auxiliary inputs. There, rather than relying on highly voluminous system calls for detection, we can potentially boost the effectiveness of a simple system that tracks only file write, execution, and process spawning, by combining the process path, action, and target, with the associated static features. Our approach could also be easily extended to a variety of other malware types and contextual sources as well as other security applications. For example, detecting cross-site-scripting attacks (XSS), where we could use textual HTML and JavaScript, the URL itself, and a rendered image of the website as separate inputs to a multi-view model. 

The LIME analysis that we conducted in Section \ref{sec:discussion} demonstrates that the multi-view model learns to distill contextual information suggestive of actual malicious/benign concepts; not merely statistical artifacts of the dataset, though as we observed, it can learn such artifacts as well. This underscores the need for data that reflects deployment use cases as well. Interestingly, techniques like LIME have applications beyond validating whether our model has learned the proper concepts. For example, in an endpoint detection and response (EDR) context, where analytic tools allow users that are not malware/forensics experts to perform some degree of threat hunting, we would like to be able to let users see suspicious file paths on disk. Doing this with a similarity comparison to known file paths could potentially reveal PII of other customers. Importance highlighting, like we illustrated in Figure \ref{fig:Posandnegurls}, is potentially a powerful PII-free alternative to the nearest neighbor approach.

One area which we plan to explore in future work is how to additionally utilize the large number of already labeled malware and benignware from intelligence feeds to boost training. Such feeds provide vastly greater malware diversity than typically exists on customer endpoints, since they rarely get infected. Unfortunately, those feeds do not provide file paths associated with actual infected files in the wild, and our current training regime assumes that both content and contextual data are always present during training. While several methods have been proposed for dealing with missing data \cite{garcia2010pattern}, it is not clear how to best apply these methods to file paths for our multi-view model.

Finally, our fixed-length convolutional embedding for file paths is not the only featurization scheme that we could employ. While the size of our dataset discourages training with recurrent neural networks due to lengthy training times, frameworks like PyTorch \cite{paszke2017automatic} trivially support different input lengths during training -- even for convolutional feed-forward models, so long as the intermediate convolution outputs are combined to a fixed dimension prior to hitting fully connected layers.

\begin{acks}
This research was funded by Sophos PLC.
\end{acks}

\newpage
\bibliographystyle{ACM-Reference-Format}
\bibliography{references}

\counterwithin{figure}{section}
\counterwithin{table}{section}
\newpage
\onecolumn
\twocolumn
\appendix

\onecolumn
\section{Appendix}

\subsection{File Path Lengths}
\label{sec:fp_lengths}

\begin{figure}[!h]
    \centering
    \includegraphics[width=0.5\linewidth]{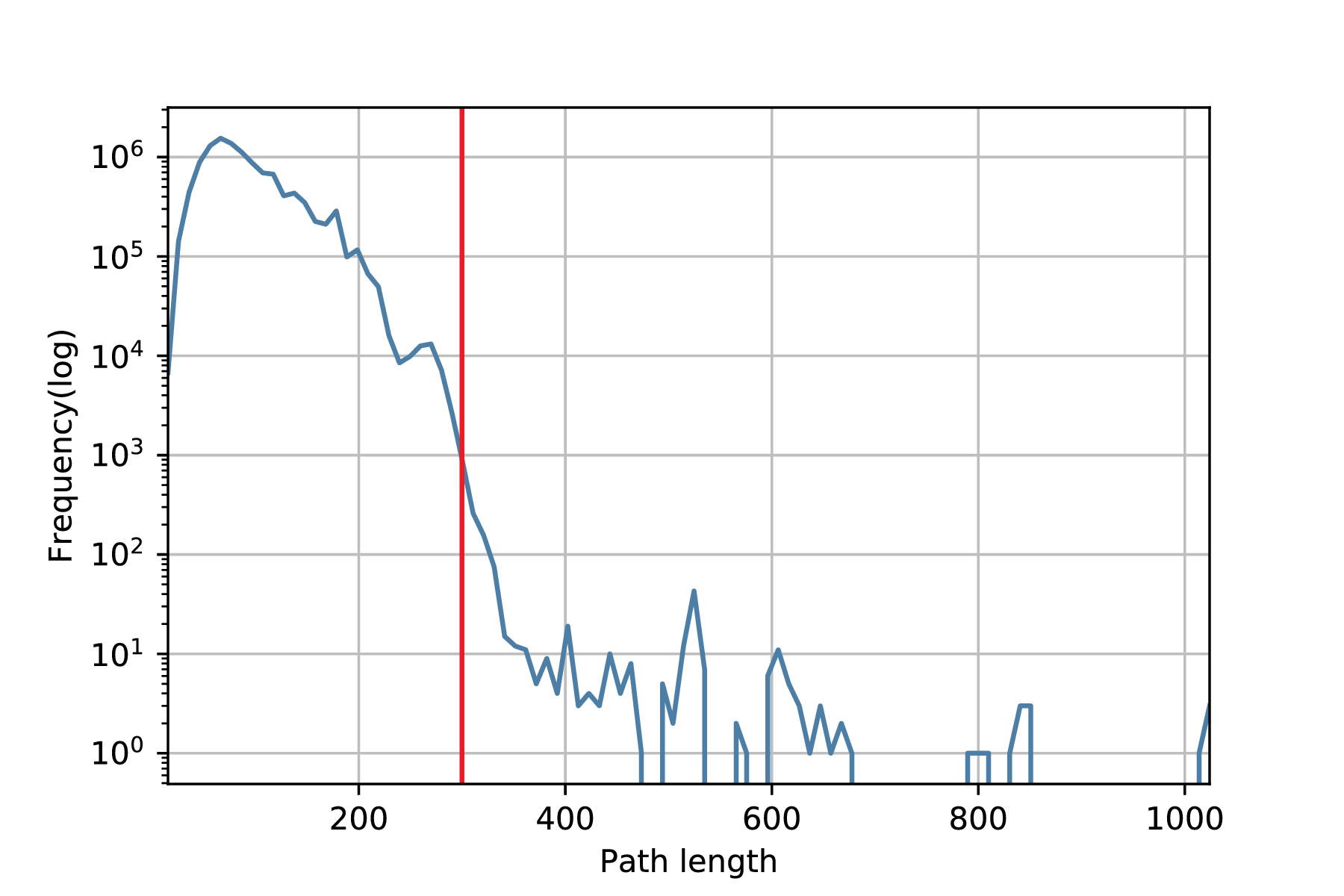}
    \caption{file path length distribution. Note the logarithmic scale on the y-axis. The vast majority of file paths are length 300 or less.}
    \label{fig:fp_length_hist}
\end{figure}

\begin{figure}[!h]

    \centering
    \includegraphics[width=0.5\linewidth]{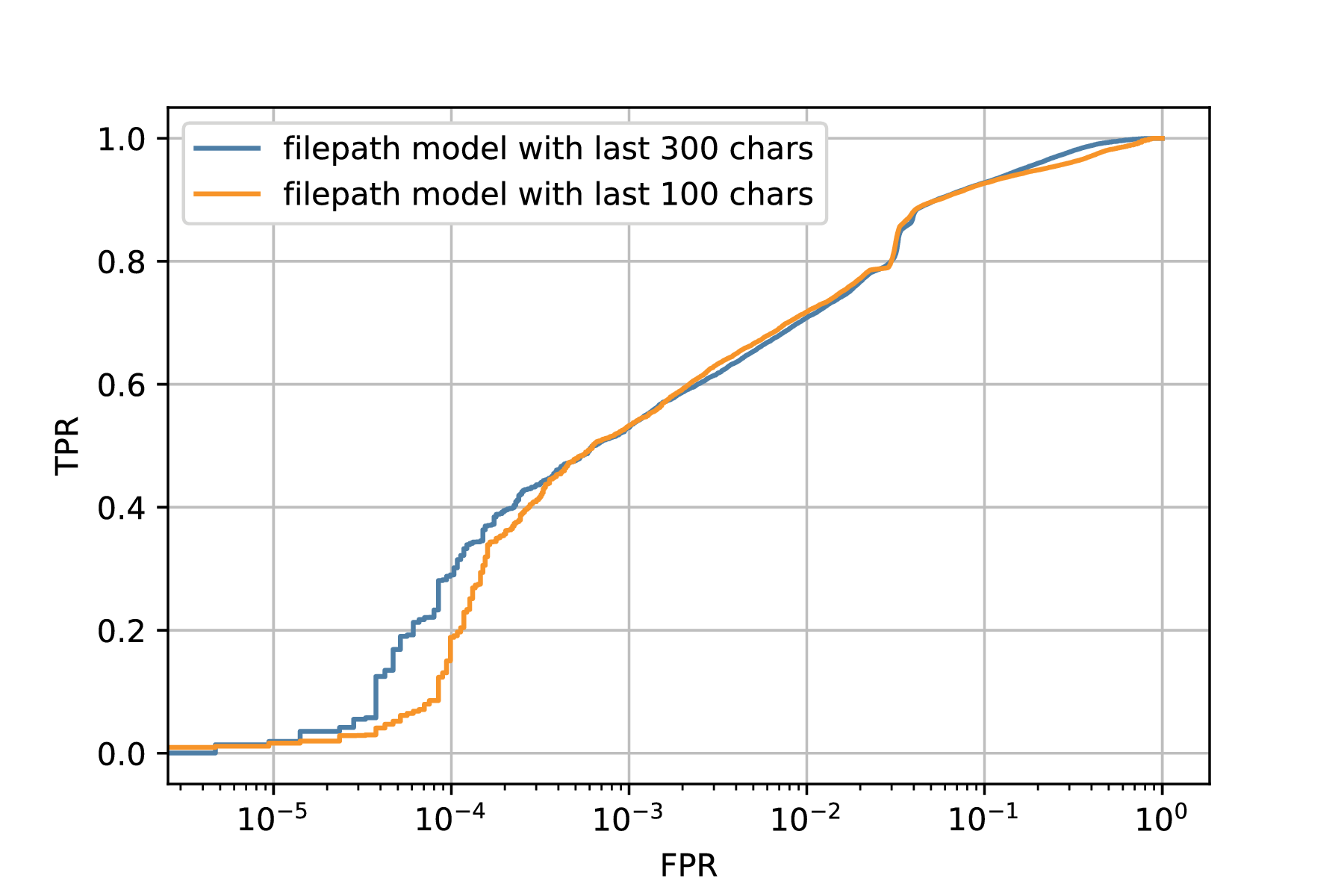}
    \caption{Performance comparison of a FP model trained using an input window size of 300 vs. an input size of 100. The length 300 model takes substantially longer to train and yields only slightly better performance.}
    \label{fig:fp_length_comparison}
\end{figure}

When selecting the input window size for our file path arm, we first examined the distribution of file path lengths from our training set. A histogram of frequencies is shown in Figure \ref{fig:fp_length_hist}. From this distribution, we initially decided to trim our file path input size to the last 300 characters. This captures the vast majority of file paths. During early experimentation, however, we found that this led to lengthy training times. In the interest of reporting uncertainty margins for each model (see Section \ref{sec:experiments}), which requires fitting multiple models, and due to our limited Amazon Web Services (AWS) budget, we first trained two models -- one that takes the last 300 characters as input and another that takes the last 100 characters as input and performed a comparison on our unfiltered test set. As shown in Figure \ref{fig:fp_length_comparison}, the performance of the length 100 model is only slightly worse than that of the length 300 model. Thus, we trimmed file paths to the last 100 characters for our experiments.  

\newpage
\subsection{LIME for the Unfiltered Test Set}
\label{sec:lime_unfiltered}

\begin{table}[!h]
    \begin{center}
    \caption{Tokens and responses from our LIME analysis selected from our unpruned dataset which \protect\subref{tab:increase_malicious} most increased the model's score for malicious samples,  \protect\subref{tab:decrease_malicious} decreased the model's score for malicious samples, \protect\subref{tab:increase_benign} increased the model's score for benign samples, \protect\subref{tab:decrease_benign} decreased the model's score for malicious samples. Characters in the middle of tokens greater than $24$ characters in length were replaced with ``...'' for readability. All of these lengthy strings contain hex digests. Malicious samples were selected from the $100$ samples from the unpruned data set where the ground truth label was malicious and the calibrated score from the PE+FP model increased the most above the calibrated score from the PE model. Benign samples were similarly selected from the unpruned data set according to the most significant drop in calibrated score. Tokens showing [xx$\hdots$xx], have been redacted for review and PII reasons.  \label{tab:tokens_weights}}
    
    \subfloat[Increase (Malicious)]{
    \begin{tabular}{lll}
    \textbf{Token} & \textbf{Weight}\\
    \hline
    \url{48bc9c40206c...bb5ebab2b3ea} & 26.115\\
    \url{1b62d0d9813d...2748ab0131a7} & 25.659\\
    \url{1764c2d644e6...60faebdb50d1} & 17.801\\
    \url{3bd39229b7ad...fc8c544b9981} & 15.978\\
    \url{25689805bb60...b1939907e6f9} & 15.967\\
    \url{d108e027c5d1...45e3820e79f2} & 15.505\\
    \url{6bae1743e31f...48f09f0bc49c} & 14.718\\
    \url{63bbd56b2099...9d9032cfad61} & 13.647\\
    \url{f72f6b477b35...c6dd77700546} & 13.639\\
    \url{ac255cc64451...7e9fa7d8a74f} & 13.292\\
    \end{tabular}\label{tab:increase_malicious}
    }
    \subfloat[Decrease (Malicious)]{
    \begin{tabular}{lll}
    \textbf{Token} & \textbf{Weight}\\
    \hline
    \url{glwnv} & -7.503\\
    \url{ghy6n} & -7.214\\
    \url{crashreports} & -6.962\\
    \url{[xxxxxxxxx]...a20d40fe645d} & -6.789\\
    \url{dll} & -5.928\\
    \url{[xxxx]engineeringtools} & -5.734\\
    \url{welivzoqf3uils} & -5.711\\
    \url{glx9azsenh1mgt} & -5.576\\
    \url{part} & -4.871\\
    \url{5iofk3xeixypt2} & -4.833\\
    \end{tabular}\label{tab:decrease_malicious}
    }\\
    \subfloat[Increase (Benign)]{
    \begin{tabular}{lll}
    \textbf{Token} & \textbf{Weight}\\
    \hline
    \url{cravingexplorer} & 13.464\\
    \url{bundled} & 9.034\\
    \url{auto~system~care} & 6.431\\
    \url{auscsetup} & 5.048\\
    \url{ubzobezzie4db} & 5.018\\
    \url{uxu5gqh} & 4.895\\
    \url{fla4476} & 4.89\\
    \url{jvuqb6pzvtju2g1} & 4.883\\
    \url{ekvzxxm} & 4.608\\
    \url{85qpk7evakbqzeo} & 4.372\\
    \end{tabular}\label{tab:increase_benign}
    }
    \subfloat[Decrease (Benign)]{
    \begin{tabular}{lll}
    \textbf{Token} & \textbf{Weight}\\
    \hline
    \url{msi8a92} & -15.498\\
    \url{robloxplayerlauncher} & -12.347\\
    \url{ultimate} & -8.59\\
    \url{debugview++} & -7.885\\
    \url{registerdll} & -7.424\\
    \url{digiarty} & -7.375\\
    \url{steganos} & -7.221\\
    \url{updates} & -7.053\\
    \url{zpsiohortfa} & -6.421\\
    \url{fslib} & -5.944\\
    \end{tabular}\label{tab:decrease_benign}
    }
\end{center}

\end{table}
\newpage

\subsection{Performance Evaluation on the Unfiltered Test Set}
\label{sec:perf_unfiltered}

\begin{figure}[!h]
    \centering
    \includegraphics[width=0.5\linewidth]{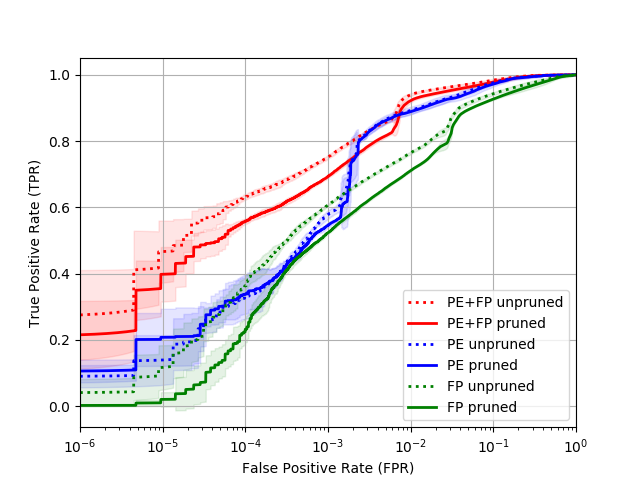}
    \caption{Mean ROC curves and standard deviations evaluating on the pruned (solid lines) and unpruned (dotted lines) test sets for comparison. PE+FP model results are shown in red. PE model results are shown in blue. FP model results are shown in green. Mean and standard deviations were evaluated over five random network initializations.
    }
    \label{fig:pruned_unpruned}
\end{figure}

\begin{table*}[!h]
\centering
\caption{Top: Mean and standard deviation true positive rates (TPRs) from evaluating our trained models on the unpruned test set at false positive rates (FPRs) of interest. Results were aggregated over five training runs with different weight initializations and minibatch orderings. Best results, shown in \textbf{bold} consistently occurred when using both feature vectors from the file and contextual file path as inputs (PE+FP). Bottom: percentage reduction in mean detection error achieved by using the the PE+FP model in comparison to the baseline content-only model (PE).\label{tab:results}}
\begin{tabular}{llllll}
\cline{2-6}

& \multicolumn{5}{|c|}{FPR}
\\\multicolumn{1}{l|}{}
&\multicolumn{1}{c|}{$10^{-5}$}
&\multicolumn{1}{c|}{$10^{-4}$}
&\multicolumn{1}{c|}{$10^{-3}$}
&\multicolumn{1}{c|}{$10^{-2}$}
&\multicolumn{1}{c|}{$10^{-1}$}
\\ \hline \hline

\multicolumn{1}{l||}{PE+FP TPR ($\mathbf{0.994}\pm0.001$ AUC)}
& \multicolumn{1}{l|}{\textbf{0.411} $\pm$ 0.072}
& \multicolumn{1}{l|}{\textbf{0.626} $\pm$ 0.006}
& \multicolumn{1}{l|}{\textbf{0.749} $\pm$ 0.003}
& \multicolumn{1}{l|}{\textbf{0.938} $\pm$ 0.003}
& \multicolumn{1}{l|}{\textbf{0.984} $\pm$ 0.005}
\\\hline

\multicolumn{1}{l||}{PE TPR \,\,\,\,\,\,\,\,\,\,($0.991 \pm 0.001$ AUC)}
& \multicolumn{1}{l|}{0.135 $\pm$ 0.063}
& \multicolumn{1}{l|}{0.304 $\pm$ 0.07}
& \multicolumn{1}{l|}{0.570 $\pm$ 0.008}
& \multicolumn{1}{l|}{0.885 $\pm$ 0.009}
& \multicolumn{1}{l|}{0.975 $\pm$ 0.007}
\\\hline

\multicolumn{1}{l||}{FP TPR \ \,\,\,\,\,\,\,\,\,($0.975 \pm 0.002$ AUC)}
& \multicolumn{1}{l|}{0.091 $\pm$ 0.063}
& \multicolumn{1}{l|}{0.365 $\pm$ 0.018}
& \multicolumn{1}{l|}{0.609 $\pm$ 0.003}
& \multicolumn{1}{l|}{0.768 $\pm$ 0.001}
& \multicolumn{1}{l|}{0.945 $\pm$ 0.002}
\\\hline\hline

\multicolumn{1}{l||}{\% Detection Error Reduction}
& \multicolumn{1}{l|}{31.9}
& \multicolumn{1}{l|}{46.3}
& \multicolumn{1}{l|}{41.7}
& \multicolumn{1}{l|}{46.1}
& \multicolumn{1}{l|}{35.5}
\\ \hline\end{tabular}

\end{table*}

\newpage
\subsection{Baseline Model Diagrams}

\begin{figure}[!h]
    \centering
    \includegraphics[width=0.75\linewidth]{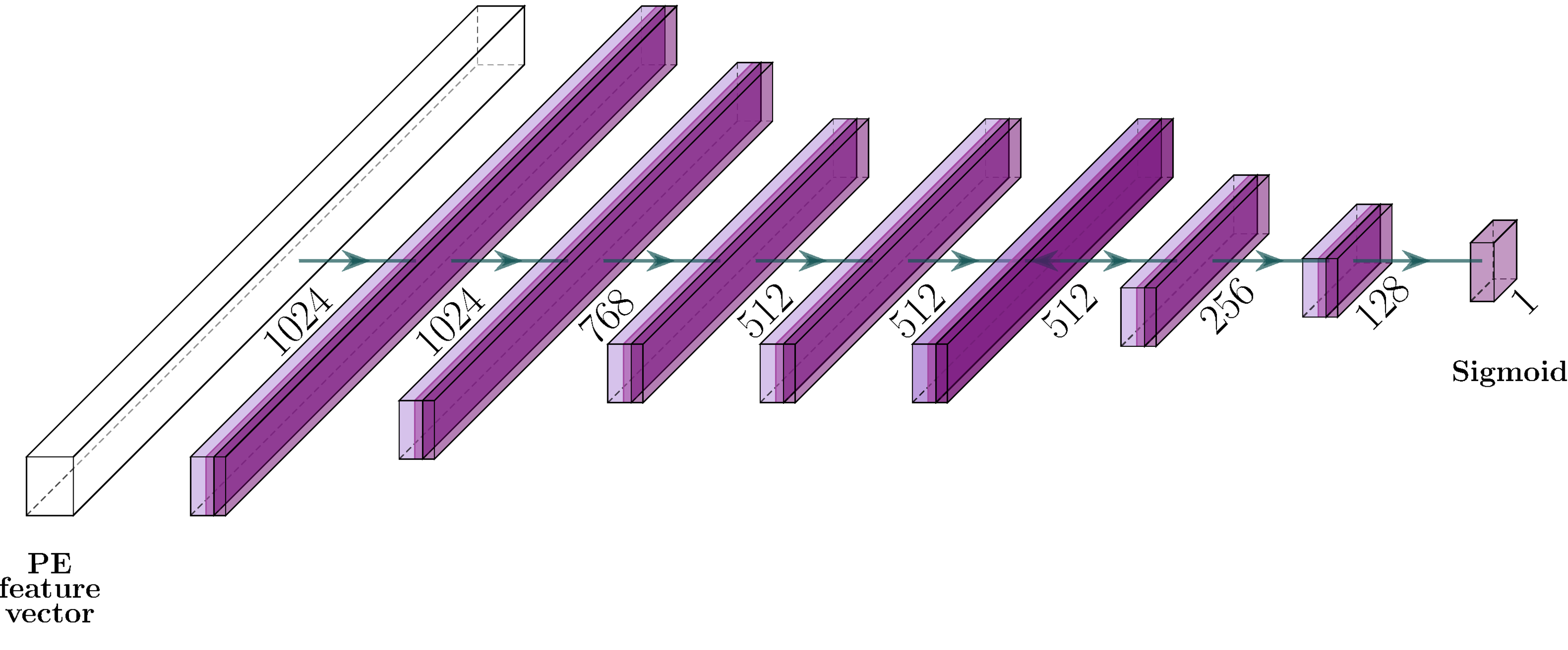}
    \caption{The PE model. Each of the unlabeled blocks contains a fully connected layer, followed by Layer Normalization and a Dropout layer. 
    }
    \label{fig:pe_model_diagram}
\end{figure}

\begin{figure}[!h]
    \centering
    \includegraphics[width=0.75\linewidth]{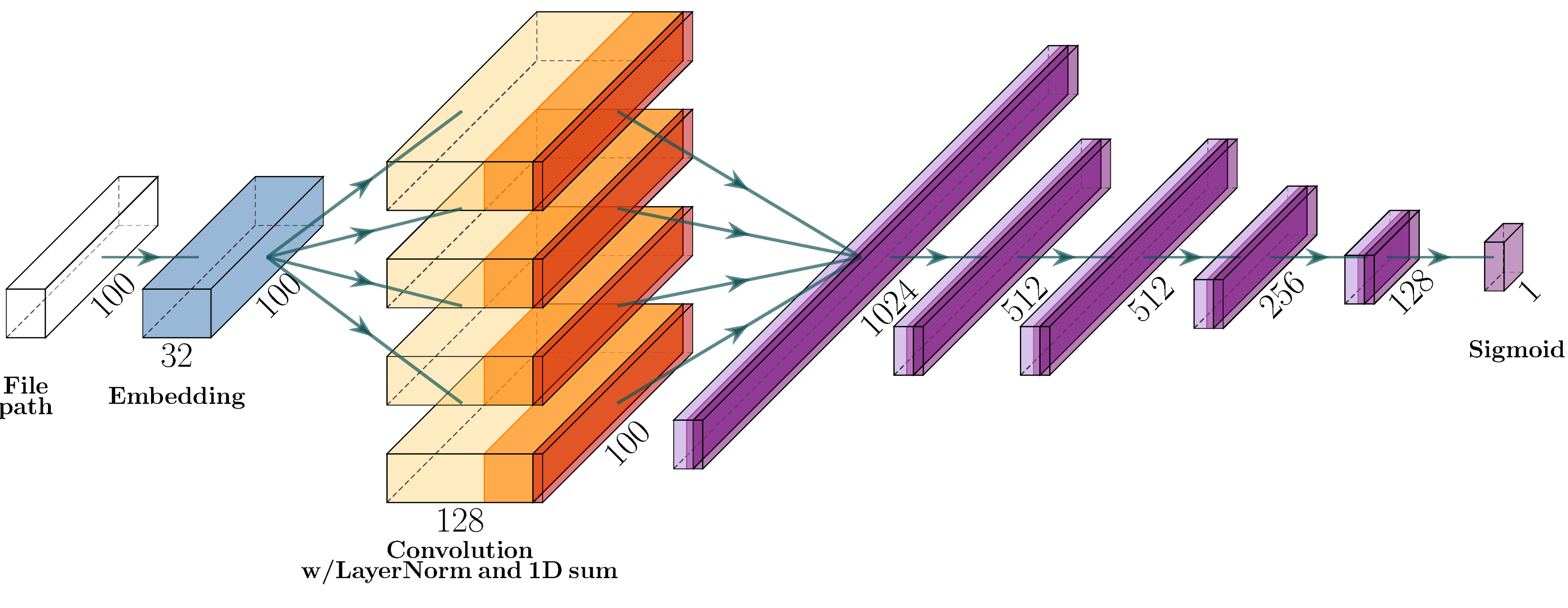}
    \caption{The FP model. Each of the unlabeled blocks contains a fully connected layer, followed by Layer Normalization and a Dropout layer. 
    }
    \label{fig:fp_model_diagram}
\end{figure}

\end{document}